\documentclass[jgrga]{agutex}
\usepackage{xcolor}
\usepackage{ulem}

\definecolor{vertf}{rgb}{0.,0.55,0.55}
\newcommand{\cv}[1]{\textcolor{black}{{#1}}}
\newcommand{\cvv}[1]{\textcolor{black}{{#1}}}
\newcommand{\cvb}[1]{\textcolor{black}{{#1}}}
\newcommand{\cvbb}[1]{\textcolor{black}{{#1}}}
\newcommand{\pv}[1]{\textcolor{black}{\sout{}}}

\usepackage[T1]{fontenc}
\usepackage[utf8]{inputenc}
\usepackage{nicefrac}
\usepackage{textcomp}
\usepackage{amsmath}
\usepackage{graphicx}

\graphicspath{{./Fig_submittedII/FigII_pdf/}}

\authorrunninghead{NEUVILLE ET AL.}

\titlerunninghead{ASPERITIES INFLUENCE ON HYDROTHERMAL EXCHANGES}

\authoraddr{E.G. Flekk{\o}y, Advanced Material and Complex System group, Department of Physics, University of Oslo, PO BOx 1048,
  Blindern, Oslo 0316, Norway. (e.g.flekkoy@fys.uio.no)}

\authoraddr{A. Neuville,
Advanced Material and Complex System group, Department of Physics, University of Oslo, PO BOx 1048,
  Blindern, Oslo 0316, Norway.
(amelie.neuville@fys.uio.no)}

\authoraddr{R. Toussaint, Institut de Physique du Globe de Strasbourg, Université de Strasbourg/EOST, CNRS, 5 rue Ren\'{e} Descartes, 67084,
  Strasbourg Cedex, France. (renaud.toussaint@unistra.fr)}

\begin{document}

%
%

\title{Influence of asperities on fluid and thermal flow in a
  fracture:  a coupled Lattice {B}oltzmann study}
%
%
%
%
\author{\large A. Neuville \altaffilmark{1}, E.G. Flekk\o y
  \altaffilmark{1,3}, and R. Toussaint \altaffilmark{2,3}}
\altaffiltext{1}{Advanced Material and Complex System group, Department of Physics, University of Oslo, PO BOx 1048,  Blindern,
  Oslo 0316, Norway.}
\altaffiltext{2}{Institut de Physique du Globe de Strasbourg, Université de Strasbourg/EOST, CNRS, 5 rue Ren\'{e} Descartes, 67084, Strasbourg Cedex, France.}
\altaffiltext{3}{The Centre for Advanced Study (CAS) at the Norwegian Academy of Science, Oslo, Norway.}








%
%


\begin{abstract}
  The characteristics of the hydro-thermal flow which occurs when a
  cold fluid is injected into a hot fractured bedrock depend on the
  morphology of the fracture.  We consider a sharp triangular
  asperity, invariant in one direction, perturbing an otherwise flat
  fracture. We investigate its influence on the macroscopic hydraulic
  transmissivity and heat transfer efficiency, at fixed low Reynolds
  number. In this study, numerical simulations are done with a coupled
  lattice Boltzmann method that solves both the complete Navier-Stokes
  and advection-diffusion equations in three dimensions. The results
  are compared with those obtained under lubrication approximations
  which rely on many hypotheses and neglect the three-dimensional (3D)
  effects. The lubrication results are obtained by analytically
  solving the Stokes equation and a two-dimensional (integrated over
  the thickness) advection-diffusion equation. We use a lattice
  Boltzmann method with a double distribution (for mass and energy
  transport) on hypercubic and cubic lattices.  Beyond some critical
  slope for the boundaries, the velocity profile is observed to be far
  from a quadratic profile in the vicinity of the sharp asperity: the
  fluid within the triangular asperity is quasi-static. We find that
  taking account of both the 3D effects and the cooling of the rock,
  are important for the thermal exchange. Neglecting these effects
  with lubrication approximations results in overestimating the heat
  exchange efficiency. The evolution of the temperature over time,
  towards steady state, also shows complex behavior: some sites
  alternately reheat and cool down several times, making it difficult
  to forecast the extracted heat.
\end{abstract}

%
%

%

\begin{article}

%
%

\section{Introduction}

Conductive and convective transport of heat or chemical species, is
omni-present in Earth sciences \citep{Stephansson04,Steefel05},
 within
porous or fractured media. Some technologies related to chemical
transport, like radioactive storage
\citep{Cvetkovic04,Amaziane08,Halecky11,Hoteit04}, or well acidizing,
requires a good resolution of advection diffusion of chemical
concentration \citep{Szymczak09, Cardenas07}. The transport is
influenced by the temperature which may play a role by modifying 1)
the fluid transport -- notably by natural convection, 2) the chemical
constants of the reactions, or 3) the geometry of the porous
medium. For instance some chemical reactions or rheological rock
transformations only occur in given ranges of temperature (e.g.
decarbonation, dehydration of sediments, decomposition of kerogen
\citep[e.g.][]{Mollo11,Petersen10}). Thermal fracturing can result
from chemical reactions generating gas \citep{Kobchenko11}, or from
thermal stress \citep{Lan12,Bergbauer98}, or from hydro-thermal
stress, when injecting hot or cold fluid into a rock. Temperature
monitoring is for example necessary to prevent potential damages of well
installations. Apart from fracturing processes, other changes of
geometry of the porous medium can also occur during injection of cold
or warm fluid in Enhanced Geothermal Systems (EGS), because of
poroelastic effects \citep{Gelet12}, and also because of chemical
reactions like acidizing. Exploitation of EGS requires also the heat
exchange to be efficient and durable. An important step is, therefore,
to understand the hydro-thermal coupling between fluid and rock; this
is the aim of our present modeling.

Hydraulic transport mostly occurs in fractures. It was shown under
lubrication approximations and steady state conditions, that the
complexity of the fracture topography influences the hydro-thermal
exchange when a cold fluid is injected into a hot fractured bedrock
\citep{Neuville10PRE}.  More specifically, the lubrication
approximations state that the aperture and its wall topographies vary
smoothly so that the velocity field is parallel to the main plane of
the fracture (the component of the hydraulic flow perpendicular to the
main fracture plane is neglected), and that the thermal diffusion only
occurs in the directions perpendicular to the main plane of the
fracture.  For this modeling, the rock temperature was also supposed
to be constant.  In other models, as e.g. in the study of
\citet{Natarajan_thermal_10}, the heat diffusion in the rock is
considered, but with a simplified fluid flow.  Some features which are
observed in nature, like fluid recirculation, and time-dependent
temperature at the pumping well, can, however, not be explained with
these model.  We therefore wish to go beyond this lubrication
assumption and be especially able to observe effects due to highly
variable morphology of the fluid-rock interface. Indeed, even if many
fluid rock interface topographies or fracture apertures are
statistically self-affine (multi-scale property)
\citep{Brown85,Bouchaud97,NeuvilleDraixI09,Candela09}, it is very
often possible to observe some isolated asperities with sharp
variations of the topography, for instance along cleavage planes
\citep{NeuvilleDraixI09}, or due to particle detachment, or along
stylolite teeth \citep{Renard04,Ebner10,Koehn12,Laronne12,Rolland12}.
The roughness of the fracture can also be perturbed by intersection
with other fractures \citep{Nenna11}, or, for microfractures, by the
matrix porosity \citep{RenardBernard09}.

\cv{Investigations on the validity of the lubrication approximation,
  based on the study of some geometrical parameters have been
  performed e.g.  by \citet{Zimmerman96, Oron98,Nicholl99}.} Without
the lubrication approximation, i.e. with full solving of the
Navier-Stokes equation, the hydraulic behavior in channels or fracture
with sinusoidal walls were studied e.g. by
\citet{BrownStock95,Miles98,Bernabe00} with lattice gas methods. It
was shown that for a sinusoid with a short wavelength and large
amplitude compared to the mean aperture, the hydraulic aperture is
smaller than that expected with the lubrication
approximation. \pv{Analytical error estimates about the hydraulic
  aperture made under lubrication approximation were provided by
  \mbox{\citet{Zimmerman96}}. }\cv{Error on the hydraulic aperture
  computed with a lubrication approximation have been analytically
  obtained by \citet{Zimmerman96}, and experimentally by
  \citet{Oron98,Nicholl99}.}  In these studies, the fluid flow was
reported to be important in the middle of the channel, while it is
quasi-stagnant in the sharp hollows of the walls. Eddies were
numerically observed in this quasi-stagnant zones
\citep{BrownStock95,Brush03, Boutt06, Cardenas07,Andrade04} in various
fracture geometries, including realistic fracture geometries. These
eddies are similar to those analytically predicted by
\citet{Moffatt64}, who studied eddies formation in a corner between
two intersecting planes, when the flow is imposed to be tangential to
the planes at an infinite distance from the corner. \cv{Microfluidics
  has also been investigated using LBM \citep[e.g.][]{Harting10}.}

On the one hand, many studies exist about the coupling between the
fully solved hydraulic behavior -- solving of the Navier Stokes
equation -- and solute or particle transport, or dispersion in general
\citep{Boutt06,Cardenas07,Drazer01,Drazer02,FlekkoyBGK93,Yeo11,Johnsen06,Niebling10,Vinningland12}.

On the other hand, few studies take into account the fully solved
hydraulic flow with the heat transfer, when a cold fluid is injected
into a fracture embedded in a hot solid.  In the absence of flow, the
stationary problem of heat transport accross a fractal interface was
studied, e.g. by \citet{Grebenkov07}. Even if both, heat and solute
transport, are described with advection-diffusion equation, many
differences exist due to different boundary conditions and different
range of parameters.  \citet{Andrade04} solved the temperature field
when a warm fluid is injected within a two-dimensional (2D) channel
delimited by walls whose topographies follow Koch fractals, using the
``Semi-Implicit Method for Pressure Linked Equations'' algorithm
developed by \citet{Patankar80}.  In his study, the walls of the
channel were however set at a constant temperature. \cv{Other studies
  regarding cooling issues of electrical devices have also been done
  \citep[e.g.][and references therein]{Young98}, but in contrast to
  the current study, the thermal boundary conditions used in these
  works are less relevant for natural problems (for instance,
  insulated walls are used)}.  Another branch of algorithms for heat
solving are done with lattice Boltzmann methods (LBM).  The LBM
\citep[e.g.][]{WolfG05} are very suitable to model the complexity of
hydrothermal transport in a rough fracture morphology, whatever the
slopes of the morphology (i.e., without any conditions on the
smoothness of the morphology). As the algorithms require only local
operations (while other numerical methods requires e.g. inversion of
matrices depending on the geometry of the entire porous medium), they
handle complex boundaries very well.  Several methods have been
proposed: multispeed scheme (a density distribution is used, with
additional speeds and high order velocity terms in the equilibrium
distribution), hybrid method (hydraulic flow is solved with LBM, and
heat transport is solved with another method), and double
distribution. A review of these methods can be found in
\citet{Lallemand03, Luan12}.  Most of the problems solved so far with
LBM deal with benchmarks that consist in close systems (square cavity
with impermeable walls).  In this paper we will examine the limits of
the lubrication approximation, and the model developed in
\citet{Neuville10PRE}. After briefly recalling the fundamentals of
this model, we describe the numerical methods used for our modeling
outside the lubrication approximations.  We chose a LBM with a double
distribution method, where the hydraulic flow is independent of the
temperature. The chosen lattices for the flow and temperature
variables are respectively hypercubic and cubic, with a single lattice
speed for each lattice. This choice is seldom used in literature, but
it is suitable for three-dimensional (3D) mass and heat transport
modeling \citep{FCHC86,WolfG05,Hiorth08}. Despite the methods being
implemented in 3D, for simplicity reasons, the parameter exploration
is done in two dimensions, translational invariance being assumed
along the third dimension. For a self-affine aperture, the
contribution of each asperity on the total hydraulic and thermal
exchanges is difficult to single out. For this reason, here, we chose
as a simpler situation to focus on one single asperity where we can
lead a precise quantitative study of the flow organization and
properties in space and time as function of flow speed, asperity size
and shape, and heat transport properties.

We will thus explore the behavior of the flow in a fracture with a
triangular asperity, as function of the asperity size. We will compare
the results directly to the lubrication approximation results, and
establish when this one fails to model correctly the mass and heat
transport.

\section{Methods for hydro-thermal modeling}
        
\subsection{Solving under lubrication approximations}
\label{sec:with-lubr-appr}
The lubrication approximation holds in the laminar regime, at small
Reynolds number, for fluids flowing into a fracture whose aperture and
both wall-topographies, show smooth variations. Under these
assumption, the Navier-Stokes equation reduces to the Reynolds
equation \citep{Pinkus61,Brown87}:
\begin{equation}
\boldsymbol{ \nabla} \cdot \left ( a^3(x,y) \boldsymbol{ \nabla} p \right )= 0, 
\label{ReynoldsEq}
\end{equation}
where $\boldsymbol{ \nabla} p$ is the local 2D pressure gradient, and
$a(x,y)$ is the fracture aperture. With $ \hat{ \boldsymbol{ x}}$ (unitary
vector) as the direction of the macroscopic pressure gradient, the
velocity expresses as
\begin{equation}
  \mathbf{u}(x,y,z)=\frac{\nabla p(x,y)}{2
  \eta}[z_1(x,y)-z][z-z_2(x,y)]\hat{ \boldsymbol{ x}},
\label{eq:speed}
\end{equation}
where $ z_1$ and $z_2$ are the out of plane coordinates of the
fracture walls related to the aperture by $z_2-z_1=a$, and $\eta$ the
dynamic viscosity of the fluid.  The aperture of a fracture $a$ is
partially characterized by its mean geometrical aperture, $A$, and by
the standard deviation of the aperture\pv{, $\sigma=\left(\langle
  a^2\rangle -A^2\right)^{1/2}$, where
$\langle.\rangle$ refers to the $x-y$ space averaging}. The hydraulic
behavior can be partially characterized by the flow across the
aperture, $\mathbf{q}$, defined as
\begin{equation}
\label{eq:debitQ}
 \mathbf{q}=\int_{a(x,y)} \mathbf{u}(x,y,z) dz.
\end{equation} 
The hydraulic aperture $H$ is classically defined
\citep{GuyonHulinPetit} from \pv{the space average of }the component of
$\mathbf{q}$ along the macroscopic gradient direction, $q_x$:
 \begin{equation}
H=\left(\langle q_x \rangle \frac{12\eta}{F}\right)^{\nicefrac{1}{3}},\label{eq:Hydraulic_Aperture2}
\end{equation}
where $F$ is the norm macroscopic pressure gradient, and
$\langle.\rangle$ refers to the $x-y$ space averaging. For parallel
plates geometry, $\mathbf{q}$ is a constant vector and $H=A$.
The thermal behavior of a fluid injected in a fracture (with a
self-affine aperture) embedded in a constantly warm rock was modeled
in \citet{Neuville10PRE, Neuville10JGI} under the so-called thermal
lubrication approximation. In their solving, several terms are
discarded in the heat equation: for instance the conduction occurs
only perpendicularly to the fracture mean plane ($z$), and the
convection is neglected along $z$. The thermal exchange balance in a
stationary regime resumes in

\begin{equation}
  \mathbf{q}\cdot\boldsymbol{\nabla}\overline{T}+2\frac{\chi_f}{a}\text{N\!u}\cdot\left(\overline{T}-T_{r}\right)=0,
\label{eq:EquadiffTbar}
\end{equation}
where \cv{$\chi_f$ is the thermal diffusivity of the fluid,}
$\text{N\!u}={\pm \left.\frac{\partial T}{\partial
      z}\right|_{z=z_{1,2}}a}/({T_{r}-\overline{T}})$ is the Nusselt
number, classically used to evaluate the thermal efficiency in
reference with the conductive heat flow, and
\begin{equation}
\overline{T}\left(x,y\right)=\frac{\int_{a}u_x\left(x,y,z\right)  T\left(x,y,z\right)dz}{q_x}\label{eq:Definition:T_2D}
\end{equation}
is a temperature averaged across the aperture, weighted by the
velocity.  For a parallel plates geometry, it can be shown, \cv{under
  assumptions over the temperature gradient}, that
$\text{N\!u}={70}/{17}$ \citep{TurcotteSchu, Neuville10PRE}.  In this method, the
temperature profile across the aperture follows a quartic law.  It was
shown \citep{Neuville10PRE} that for a self-affine aperture, the
averaged temperature law over the $y$ direction,
$\overline{\overline{T}}$, defined as

\begin{equation}
  \overline{\overline{T}}(x)=\frac{\int
    q_{x}\left(x,y\right)\overline{T}\left(x,y\right)/a(x,y)dy}{\int q_{x}\left(x,y\right)/a(x,y)dy},\label{Definition:T_1D}\end{equation}

 can be approximated by
\begin{equation}
\overline{\overline{T}}-T_{r}=\left({T}_f^{0}-T_{r}\right)\exp\left(-\frac{x}{R}\right),\label{eq:ExpTPlaques}\end{equation}
where ${T}_f^0=\overline{T}(x=0)$ is imposed at the inlet of the
fracture, $T_r$ is the temperature of the wall (interface
fluid-solid), and $R$ is a thermal length, obtained from a linear
regression. For parallel plates separated by $A_0$, $R$ is equal to
\begin{equation}
R=\frac{A_0 q}{2 \text{N\!u} \chi_f}.\label{eq:LengthThermalizationParallelPlates}
\end{equation}
Note that any change of the thermal length can also be interpreted in
term of Nusselt number variation.
\label{sec:char-lengths-defin}

\subsection{Full solving, using coupled Lattice Boltzmann Methods
  (LBM)}

\subsubsection{Solving the mass transport with FCHC LBM}

\cvv{
In LBM, fictitious particles move and collide on a lattice. Operations
conserves mass and momentum at mesoscopic scale.  Using appropriate
rules and lattice topology, it can be shown that the Navier-Stokes
equation can be recovered at macroscopic scale
\citep[e.g.][]{RothZaleski,Chopard98,WolfG05}. The distribution of
mass density is denoted as $f_i$, where the index $i$ refers to the
direction of propagation of the particles moving with a velocity
$\boldsymbol{c_i}$ on the lattice. We choose a
``Face-Centered-Hyper-Cubic'' (FCHC) lattice. This is a
four-dimensional lattice with suitable topological properties to
solve the Navier-Stokes equation \citep{FCHC86} in three
dimensions. The $N=24$ vectors defining the lattice directions are
$\boldsymbol{c_i}\frac{\delta t}{\delta x}$ $(\pm 1, \pm1, 0, 0)$,
$(0, \pm 1, \pm1, 0)$, $(0, 0, \pm 1, \pm1)$, $(0, \pm 1, 0, \pm1)$,
$(\pm 1, 0, \pm1, 0)$, $(\pm 1, 0, 0, \pm1)$, with $c=\sqrt 2 \delta x
/ \delta t$, where $\delta t$ and $\delta x$ are the time and space
steps.
The total density and the macroscopic velocity $\mathbf{u}$ at each
node $M$ are computed with ($\mathbf{O\!M}$ being the position
  vector):
 \begin{equation}
 \begin{split}
\label{eq:rho_sum_f}
\rho(t,\mathbf{O\!M})&=\sum_{i=1}^{24}{f_i}(t,\mathbf{O\!M})\\
\rho\mathbf{u}(t,\mathbf{O\!M})&=\sum_{i=1}^{24}{\boldsymbol{c_i}f_i}(t,\mathbf{O\!M}).
\end{split}
\end{equation}}

\cvv{For the collision phase, a standard BGK scheme \citep[Bhatnagar,
Gross, and Krook -- ][]{Bhatnagar54,Qian92} is used. The linearized
collision term depends on a constant of relaxation $\lambda$, which is
linked to $\nu=\eta/\rho$ the kinematic viscosity of the fluid
\citep{RothZaleski} by:
 \begin{equation}
\nu=-\frac{c^2\delta t}{6}\left ( -\frac{1}{2}+{\lambda} \right).
\end{equation}
}
\cvv{The macroscopic pressure gradient between the inlet and outlet of the
fracture is implemented through a volumetric force that intervenes
during the collision phase at each lattice
node.} 
 
\cvv{It can be shown that the density of the fluid is linked to the
  velocity by $p= \rho(\delta x^2/(2 \delta t)^2-u^2/4)$
  \citep{RothZaleski}. This modeling of Navier-Stokes equation holds
  for compressible flows in the incompressibility limit, at small Mach
  number (ratio between the velocity and the sound speed on the
  lattice -- here, equal to $u \delta t/(\sqrt 2 \delta x)$). The Knudsen
  number (ratio between the average distance between two collision,
  and the macroscopic scale of the system) should also be small.
\label{sec:Artefact_rho_u}}
%
%

\subsubsection{Solving the heat transport with 3D cubic LBM}

\cvv{The transport of the heat is solved using a coupled lattice
  Boltzmann method, using a second particle distribution, $g_i$, which
  represents the internal energy distribution of the particles moving
  with the velocity $\mathbf b_i$. It is assumed that the heat is
  passively transported by the fluid: viscous heat dissipation is
  neglected, as well as the viscosity dependence with the
  temperature. Using square lattices and appropriate collision rules,
  it has been shown \citep{WolfG05,Hiorth08} that the LBM solves
  advection-diffusion equation. The internal energy and its flux are
  conserved during the collision phase, which is done with a BGK
  scheme as in e.g.  \citet{Hiorth08} is used. We choose a 3D cubic
lattice. Its $N_T=6$ base vectors, $\boldsymbol{b_i}\frac{\delta
  t}{\delta x}$, are defined by $(\pm 1, 0, 0)$, $(0, \pm 1, 0)$, $(0,
0, \pm1)$ and $b=\delta x / \delta t $. This network is a sublattice
of the 3D projection -- perpendicularly to fourth direction -- of the
FCHC lattice.  The temperature at each node $M$, and the internal
energy flux are given by:}
\begin{equation}
  \label{eq:En_flux_LB_sum_gi}
  \begin{split}
    \sum_{i}^{N_T}g_{i}(\mathbf{O\!M},t)&=
    T(\mathbf{O\!M},t)\\
    \sum_{i}^{N_T}\mathbf{b}_{i}g_{i}(\mathbf{O\!M},t)&=
    T(\mathbf{O\!M},t)\mathbf{u}
  \end{split}
\end{equation}
%

\subsubsection{LBM boundary conditions}
Let consider a fractured medium, where the macroscopic pressure
gradient in the fluid is aligned with the $\hat{ \boldsymbol{ x}}$
direction: the fluid injected at the inlet ($x=0$), and pumped out at
the outlet of the system ($x=L_x$). The unitary vectors
$\hat{\boldsymbol{ x}}$, $\hat{ \boldsymbol{ y}}$ and $\hat
{\boldsymbol{ z}}$ define an orthonormal frame, and this porous medium
is discretized with a cubic mesh. The center of each mesh is a fluid
or a solid node.

\begin{figure}[thbp!]
\begin{center}
\noindent\includegraphics[width=6cm]{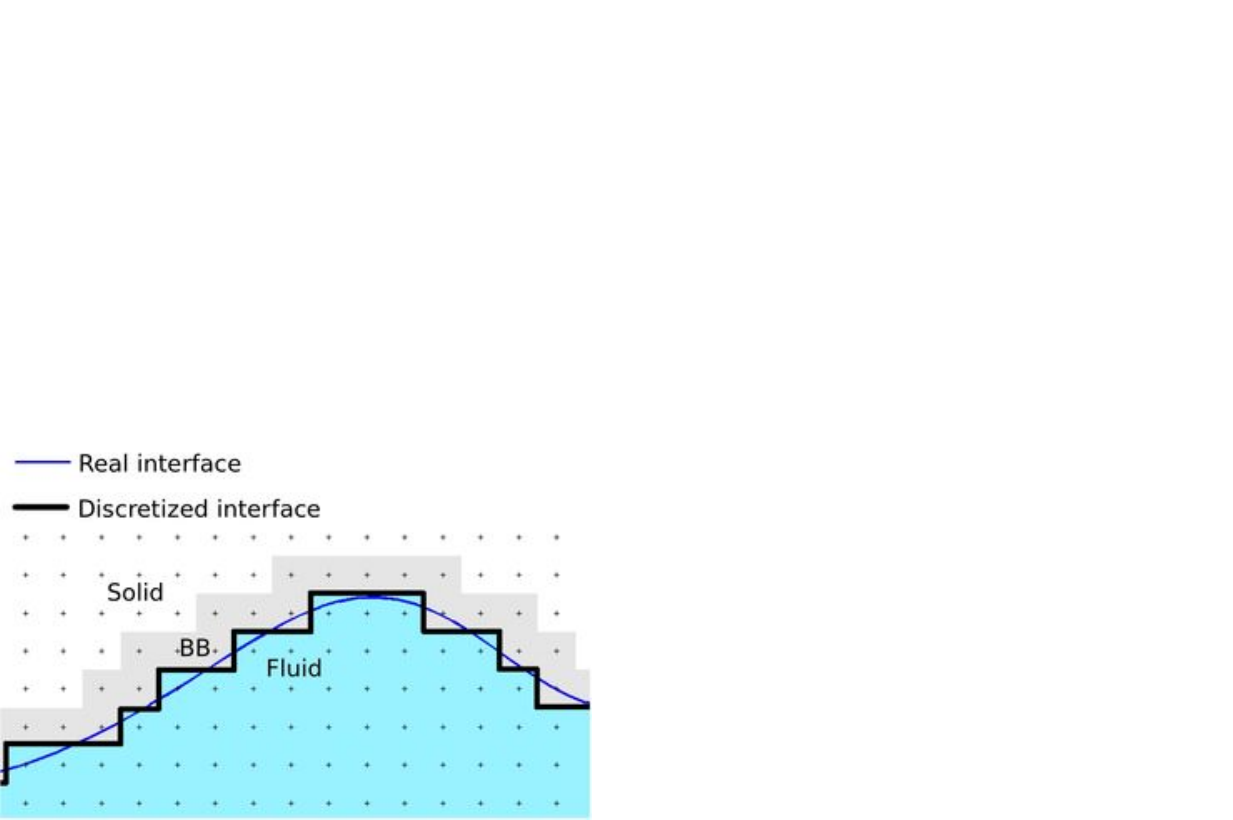}
\caption{Discretization of the interface between the solid and
  fluid. The dots indicate the nodes of the lattice. BB stands for
  nodes in light gray where some particle distributions are bounce
  backed.\label{fig:Mask}}
\end{center}
\end{figure}

The porous medium geometry and the flow are supposed to be periodic in
$x$ and $y$. At the interface between the fluid and the solid, non
slip boundary condition are chosen. \pv{In lattice-Boltzmann methods,
  t}This is implemented using the \pv{classical }full-way bounce back
boundary condition \citep{RothZaleski} for the mass particle
distributions. This bounce back operation is done for particle
distributions $f_i(\mathbf{O\!M})$ where $M$ is a solid node, and
$\mathbf{O\!M}+\boldsymbol{c_i}\delta t$ is in the fluid.  For these
nodes, \pv{instead of performing a collision operation, }the mass
distributions $f_i$ of the bounce-backed nodes are exchanged with
$f_{i+12}$, where the direction $i+12$ is opposite to the direction
$i$. \cvbb{This is done instead of a collision operation.} The
propagation phase is then done normally on these nodes. For this
bounce-back boundary condition, the interface fluid solid is supposed
to be located half-way between the bounce-backed node and the fluid
nodes (Fig.~\ref{fig:Mask}).

The temperature field is supposed to be periodic along $y$ direction.
The temperature is imposed in $x=0$, $z=0$, $z=L_z$, where $L_z$ is
the height of our system in the $z$ direction. In this study, the rock
is maintained at temperature $T_r^0=150\text{\textdegree{}C}$ at the
borders in $z=L_z$ and $z=0$. At the inlet of the system, in $x=0$,
the rock and fluid are maintained respectively at $T=T_r^0$, and
$T=T_{f}^0=30\text{\textdegree{}C}$.  At the outlet of the system, the
temperature is forced to $T=T_r^0$ at solid nodes (Dirichlet
condition), and $T(x=L_x)=T(x=L_x-\delta x)$ in liquid (von Neumann
condition).  In our program, the temperature at these nodes is imposed
through an ``equilibrium scheme'' \citep{Huang11}: at the end of each
time step, and at the next collision step, the equilibrium
distributions $g_i$ of these boundary nodes is set at the next
collision step to \cvv{the equilibrium distribution} $g_i^{\text{eq}}$,
\pv{computed from Eq.~(\ref{eq:3})}with the desired temperature. For
the nodes located at the outlet in the fluid, our boundary condition
is equivalent, at first order, to zero temperature gradient along the
$x$ direction.  The systems considered are long enough so that the
beginning of the systems is almost not influenced by outlet conditions
(this boundary condition only creates a local artifact at the outlet).
At the initial time, the rock and fluid have a temperature of respectively
$T_r^0$ and $T_f^0$.

\section{Geometry \pv{studied }and hydro-thermal regime studied}
We focus on the hydrothermal behavior within a fracture with a very
simple geometry: it is a fracture with flat walls parallel to the
$x-y$ plane, perturbed by a single asperity with sharp edges
(Fig.~\ref{fig:velocity_h20_L50}).  The fracture aperture $a(x)$ is
invariable along the $y$ direction. In cross-view ($x-z$ plane), the
asperity has a triangle shape characterized by its width $L$, depth
$d$, and abscissa position $x_0$:
\begin{equation}
  a(x)=A_0+d\:\Lambda\left[\!\!\;\!\frac{2}{L}\left(x-x_0-L/2\right)\right],
\label{eq:aperture}
\end{equation}
where $\Lambda$ is the unitary triangle function:
\begin{equation}
\Lambda(x)=
\begin{cases}
 0, &\left|x\right|\ge 1\\
1-\left|x\right|,&  \left|x\right|<1.
\end{cases}
\end{equation}

For all the computations done with LBM in this study, the fracture is
embedded in a solid whose dimension are $(L_x, L_z)=(200\text{~mm},
89.5\text{~mm})$. \cvbb{This medium}\pv{full system} can be
\cvbb{completely} seen in Fig.~\ref{fig:TemperatureMap}a.  The bottom
wall of the fracture intersects the $x-z$ plane in
$z=39.75\text{~mm}$, and the asperity is characterized by $(A_0,
x_0)=(10\text{~mm}, 5\text{~mm})$, while $d$ and $L$ vary. The lattice
discretization is $\delta x =0.5\text{~mm}$ and $\delta t =
0.1250\text{~s}$.

The LBM simulations are done at low Reynolds number:
$\text{R\!e}=0.17$, where it is computed as $\text{R\!e}={2 A
  u_M}/({3\nu})$ with $u_M$ being the maximum velocity within a flat
fracture separated by two parallel plates, of aperture $A_0$, estimated
from the classical cubic law $u_M=F A_0^2/(8\eta)$
\citep{GuyonHulinPetit}.

For the thermal parameters, two different thermal diffusivity values
are used in LBM for the fluid and the solid. The ratio of both,
$\chi_r/\chi_f=0.17$, corresponds to the typical ratio of diffusivity
values for crystalline rocks (of order of $1\text{~mm}^2$/s
\citep{Drury87}) and water (at $100\text{\textdegree{}C}$,
$0.17\text{~mm}^2$/s -- \citep{Taine03}). The Péclet number, defined
as $\text{P\!e}=u_M A/\chi_f$ is set to 45.96.\label{RePe} The orders
of magnitude of Reynolds and Péclet numbers that we use are compatible
with the lubrication approximations.

\section{Results: example of application}

\subsection{Illustration of the hydraulic behavior}

\subsubsection{Hydraulic lubrication approximation for a triangular asperity}
 The lubrication velocity is computed using
Eq.~(\ref{eq:speed}). For an aperture which is invariant
along $y$, the Reynolds equation Eq.~(\ref{ReynoldsEq}) simplifies to
\begin{equation}
\partial_x p=K/a(x)^3,
\label{eq:dp}
\end{equation}
and \cvbb{the hydraulic flow is constant:}
\begin{equation}
  \label{eq:q_lubri_geom_invariante}
  \mathbf{q}=-K/(12\eta)\hat{ \boldsymbol{ x}},
\end{equation}
 where $K$ is a constant.  By integrating the
pressure gradient between the inlet and outlet of the fracture, one
gets
\begin{equation}
K=-{F L_x}\left({\int_0^{L_x}a^{-3}(x)dx}\right)^{-1},
\label{eq:Kval}
\end{equation}
where $\boldsymbol{F}$ is the pressure gradient imposed between the inlet
and outlet of the fracture.
$K$ is calculated analytically for the considered geometry
(Eq.~(\ref{eq:aperture})), by noticing that:
\begin{equation}
\int_0^{L_x}\frac{dx}{a^{3}(x)}=\frac{-3A_0 d L - 2 d^2 L + 2 A_0^2 L_x + 4 A_0 d L_x + 2
d^2 L_x}{2 A_0^3 (A_0 + d)^2}.\label{eq:Kval_integrale}
\end{equation}
\subsubsection{Fully resolved hydraulic behavior compared to the
  lubrication approximation}

\begin{figure*}[thbp!]
\noindent\includegraphics[width=0.95\linewidth]{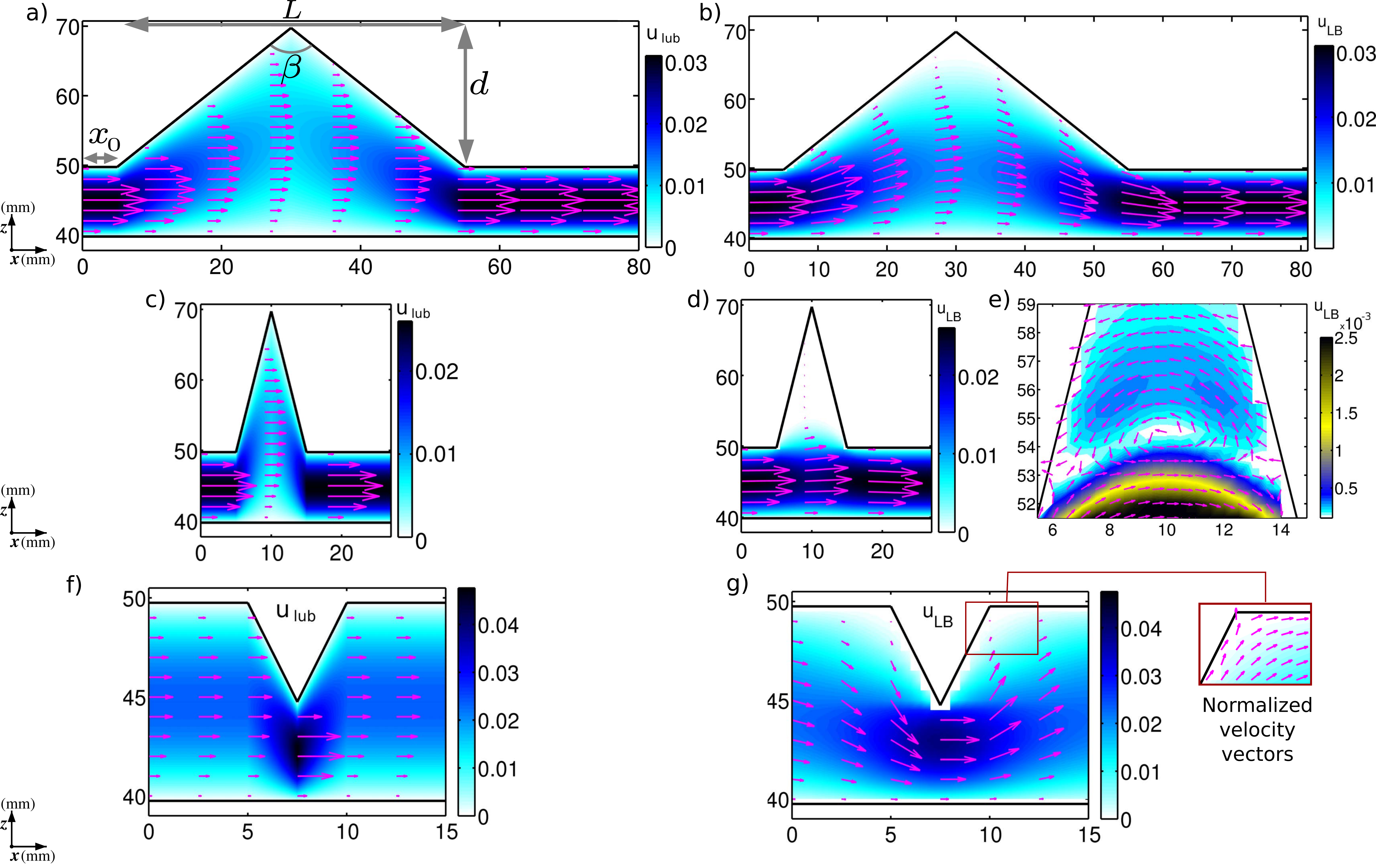}
\begin{center}
  \caption{(Color
    online) \label{fig:velocity_h20_L50}\label{fig:velocity_h20_L10}(a)
    Cross section view of the velocity norm (unit of the color scale:
    mm/s) under lubrification approximation ($u=u_x$). The shape of the asperity is
    defined by $(d, L)=(20\text{~mm}, 50\text{~mm})$. (b)  Velocity
    vectors and their norm in color, with LBM. Same color scale as Fig.~\ref{fig:velocity_h20_L50}a.
  (c, d) Same as (a, b) for $(d, L)=(20\text{~mm},
  10\text{~mm})$. (e) Zoom of (d) with a different color scale, and
  the normalized velocity vectors
  $\mathbf{u}(x,z)/\left\|u\right\|$ superimposed. \cvv{(f, g) Same as (c,
  d, e) for a bump asperity:  $(d, L)=(-5 \text{~mm}, 5\text{~mm})$. The scaling
  used to represent the vectors is six times smaller for (f, g) than for
  (c, d) and (a, b).}
 \label{fig:velocity_h_5_L5}
}
\end{center}
\end{figure*}

\cvv{Let's first comment on the precision of the lattice Boltzmann
  (LB) results.} We have some errors that come from the chosen
implementation of the boundary conditions combined to the type of
LBM. \label{error_hydrau} Because the fluid is slightly compressible
with LBM, the averaged flow $q_x$ is not constant (the relative
standard deviation of $q_x$ is around 0.39\% for this example), and
therefore the hydraulic aperture slightly varies according to the
$x-$domain where it is computed. We chose to compute $H$ at the
asperity scale, {\it i.e.}  for $x_0 \leq x \leq x_0+L$. For the case
of a parallel flat wall fracture separated by $A_0=10\text{~mm}$, with
a flow characterized by $\text{R\!e}=0.17$, solved with steps $\delta
x=0.5\text{~mm}$, and $\delta t=0.125\text{~s}$, the absolute error on
the computed velocity, defined as $E_a=\langle
\left(u_x^{\text{L\!B}}-u_x^l) \right)^2 \rangle^{1/2}$ is
$5.36\,10^{-4}\text{~mm/s}$. The relative error, defined as
$E_r=E_a/\langle u_x^l \rangle$ is 3.22\%. The relative error on the
hydraulic aperture computed in this case is 1.11\%. In order to take
into account this numerical error, the comparison between LB and
lubrication results is done using normalized hydraulic apertures. The
hydraulic aperture obtained from the LBM calculation, and the one
obtained from the lubrication approximation are respectively
normalized in this way
$H_{\text{L\!B}}^*=H_{\text{L\!B}}/H^{/\!/}_{\text{L\!B}}$ and
$H_{\text{l\!u\!b}}^*=H_{\text{l\!u\!b}}/H^{/\!/}_{\text{l\!u\!b}}$\label{def_HLBetoile}. $H^{/\!/}_{\text{L\!B}}$
and $H^{/\!/}_{\text{l\!u\!b}}$ are the hydraulic apertures in
parallel plate geometry,\label{def_ratio_H_Hlubri} computed
respectively with the LBM (discretized aperture), and with the
lubrication approximation.  We note also an error on the direction of
the velocity vectors in the deepest part of the corner, for velocity
vectors whose norm are of order $10^{-5}$~mm/s, i.e. very low velocity
compared to the average velocity (see
Fig.~\ref{fig:velocity_h20_L50}).  In the zones where artefacts at
very low velocities were observed, the thermal exchange is mainly led
by diffusive exchanges. Therefore, we estimate that the error in the
direction does not influence much the thermal exchange. Note that
lattice Boltzmann methods with better precision are also available,
like those \cvv{with a modified equilibrium distribution
  \citep{HeLuo97}, or those} with two relaxation times
\citep[e.g.][]{Talon12}.

\begin{figure*}[thbp!]
\noindent\includegraphics[width=\linewidth]{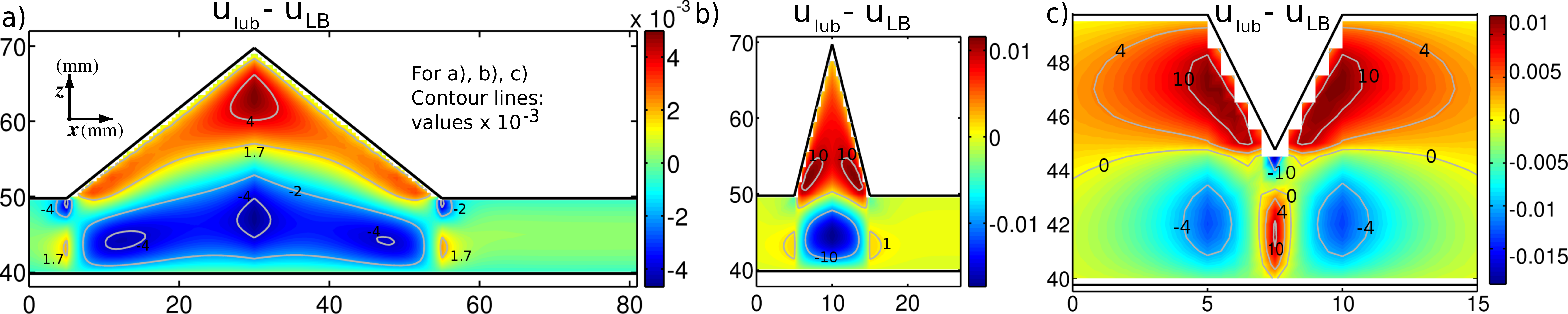}
\begin{center}
  \caption{\cv{(Color
    online) Cross section view of  the difference of the velocity
    norms obtained with
    lubrication and with LBM, for the geometries shown in Fig.~\ref{fig:velocity_h_5_L5}}
 \label{fig:velocity_diff}
}
\end{center}
\end{figure*}

Figure~\ref{fig:velocity_h20_L50}a shows the velocity norm under
lubrication approximation (cross-section view), for a fracture with an
asperity characterized by $(d,L)=(20\text{~mm},50\text{~mm})$. This
has to be compared with Fig.~\ref{fig:velocity_h20_L50}b which shows
the velocity norm and vectors at steady state across the fracture
aperture, computed with the LBM. \cv{The difference of the velocity
  norms is in addition shown in Fig.~\ref{fig:velocity_diff}a.} In
this configuration, the fluid flows within the asperity, and the main
flow direction changes gently in accordance with the topography of the
walls.  The velocity field fully resolved and the one solved with the
lubrication approximation show some similarities. However, some
details are not captured with the lubrication approximation, notably
in the deepest part of the corner where the full computation
\cvv{locally} shows fluid at rest.  It is computed that
$H_{\text{L\!B}}^*= 1.06$ and $H_{\text{l\!u\!b}}^*=1.07$ for the
geometry shown in Fig.~\ref{fig:velocity_h20_L50}a, with
$(d,L)=(20\text{~mm},50\text{~mm})$.  Those values only differs by
1.15\% i.e. the lubrication approximation still holds \cvv{on
  average}.

Let depart further from the smooth geometry where the lubrication
assumptions apply. The same geometry as previously is used, but $L$ is
set to $L=10\text{~mm}$, so that the geometry has a steeper
topography. Similarly to Figs.~\ref{fig:velocity_h20_L50}a-b,
Figs.~\ref{fig:velocity_h20_L10}c-d respectively show the lubrication
and fully resolved velocity fields. Here, it is very clear that both
are very different \cv{(see also Fig.~\ref{fig:velocity_diff}b)}.  In the asperity, a separation zone is observed:
the fluid velocity is very small, and Fig.~\ref{fig:velocity_h20_L10}e
shows that the fluid recirculates as if being trapped. The velocity
profile is consequently very different from a quadratic profile as in
Eq.~(\ref{eq:speed}).  For the geometry shown in
Figs.~\ref{fig:velocity_h20_L10}c-d $(d,L)=(20\text{~mm}, 10\text{~mm})$, it is
computed that $H_{\text{l\!u\!b}}^*=1.01$ and $H_{\text{L\!B}}^*= 1.00$.
Therefore, it means that macroscopically the lubrication approximation
is still a good estimation is this case. It is however clear that
locally, within the asperity, the lubrication approximation is not
valid.

Fractures with a bump (asperity with $d<0$) that reduces the aperture,
are also investigated. Figs.~\ref{fig:velocity_h_5_L5}f-g show the
lubrication, and fully resolved velocity fields. The main differences
are the two small separation zones \cvv{with low velocities} that
appear just before and after the bump
\cv{(Fig.~\ref{fig:velocity_diff}c)}, in the corners with obtuse
angles. This tiny asperity \pv{consequently }reduces the hydraulic
aperture of 3.0\%, with $H_{\text{L\!B}}^*= 0.96$ and
$H_{\text{l\!u\!b}}^*=0.98$.

\subsection{Illustration of the thermal behavior}
\subsubsection{Thermal lubrication approximation for a triangular
  asperity}
\label{sec:therm-lubr-appr}
 
For an aperture which is invariant along $y$, the hydraulic flow
$\mathbf{q}$ is constant (Eq.~(\ref{eq:q_lubri_geom_invariante})), and
$\overline{\overline{T}}=\overline{T}$. By assuming that the rock
temperature $T_r$ is constant, Eq.~(\ref{eq:EquadiffTbar}) simplifies
into a first order linear ordinary differential equations with
constant coefficients which has for solution (in a stationary regime):
\begin{equation}
  \label{eq:Sol_Tbar_integrale}
  {\overline{T}}-T_{r}=\left({T}_f^{0}-T_{r}\right)\exp\left(-\int^x_0\frac{A_0}{a(\xi)R}d\xi \right).
\end{equation}
For the considered geometry (Eq.~(\ref{eq:aperture})),
$\int^x_0({a(\xi)})^{-1}d\xi$ can be computed analytically and
expressed as a function of $R$
(Eq.~(\ref{eq:LengthThermalizationParallelPlates})). The solution
$\overline{T}^*=(\overline{T}-T_r)/({T}_f^{0}-T_{r})$, is:

\begin{equation}
\begin{cases}
\label{eq:Sol_Tbar_explicit}
 e^{-\frac{x}{R}},&0\le x\le x_0\\
e^{-\frac{x_0}{R}} \left[1+\frac{2d(x-x_0)}{A_0L}\right]^{-\frac{LA_0}{2dR}},& x_0\le
 x \le x_0+\frac{L}{2}\\
e^{-\frac{x_0}{R}}\left[\frac{A_0(A_0L+2d(L-x+x_0))}{L(A_0+d)^2}\right]^{\frac{LA_0}{2dR}},
 & x_0+\frac{L}{2} \le x \le x_0+L\\
e^{\frac{L-x}{R}} \left(\frac{A_0}{A_0+d}\right)^{\frac{LA_0}{dR}},& x_0+L \le x \le L_x.
\end{cases}
\end{equation}
This solution is shown in \cvbb{plot (a) of Fig.~\ref{fig:Tbar_ter}
  and (a', b', c') of Fig.~\ref{fig:Tbar_ter_b},} where
$\ln\overline{T}^*$ as a function of $x$ is plotted for several
geometries. Within the lubrication approximation, the slope of
$\ln\overline{T}^*$ as a function of $x$
(Eq.~(\ref{eq:Sol_Tbar_explicit})) is the same before and after the
asperity, \cvv{(i.e. for $x\leq x_0$ and for $x \geq x_0+L$)} and it
is given by $1/R$, where $R$ is defined in the lubrication regime with
Eq.~(\ref{eq:LengthThermalizationParallelPlates})\cvv{, $q$ being
  computed from Eqs.~(\ref{eq:q_lubri_geom_invariante}) to
  (\ref{eq:Kval_integrale})}.  Both straight lines have however
different ordinates at the origin. This comes from the complicated
behavior within the asperity zone. As a consequence the fit of
$\ln\overline{T}^*(x)$ with a single straight line, following
Eq.~(\ref{eq:ExpTPlaques}), clearly does not capture the details of
the thermal exchange. \pv{In Fig.~\ref{fig:Tbar_ter}e, }\cvbb{In
  Fig.~\ref{fig:Tbar_ter_b}e, } the linear fits done for the
restricted range $0 \leq x \leq 55$~mm \pv{is}\cvbb{(in the vicinity
  of the inlet) are} shown. The negative inverse of the slope of these
fits is named $R_{1\,\text{l\!u\!b}}$\label{def_R1lub} \cvbb{(reported
  values in Tab.~\ref{tab:Lref})}, and it can be compared to the $R$
values. For \cvbb{hollow asperities} ($d>0$),
$R_{1\,\text{l\!u\!b}}>R$. It means that, according to this solution,
the heat exchange efficiency is reduced around the corner, compared to
the heat exchange efficiency far from the corner. On the contrary, for
\cvbb{bumps} $d<0$, $R_{1\,\text{l\!u\!b}}<R$.  The linear fit of the
second part (far from the inlet), for $55 \leq x \leq 150$~mm, results
in a thermal length $R_{2\,\text{l\!u\!b}}$ \cvv{analytically} always
equal to $R$\cvv{. Indeed, since} $x_0$ and $L$ values fulfill, by
choice, $x_0+L \leq 55$~mm\cvv{, this second fit is systematically
  done after the asperity}.
\subsubsection{Fully resolved thermal behavior compared to the
    lubrication approximation}

\cvb{In this subsection, we first comment qualitatively on the maps of
  the temperature obtained with LBM for few different asperities, in
  a stationary regime. The comparison to the lubrication approximation
  is then done quantitatively using average temperatures curves, from
  which  thermal lengths are computed. The transient regime is also
  discussed at the end of this subsection.}

\begin{figure}[thbp!]
\begin{center}
\noindent\includegraphics[width=8.5cm]{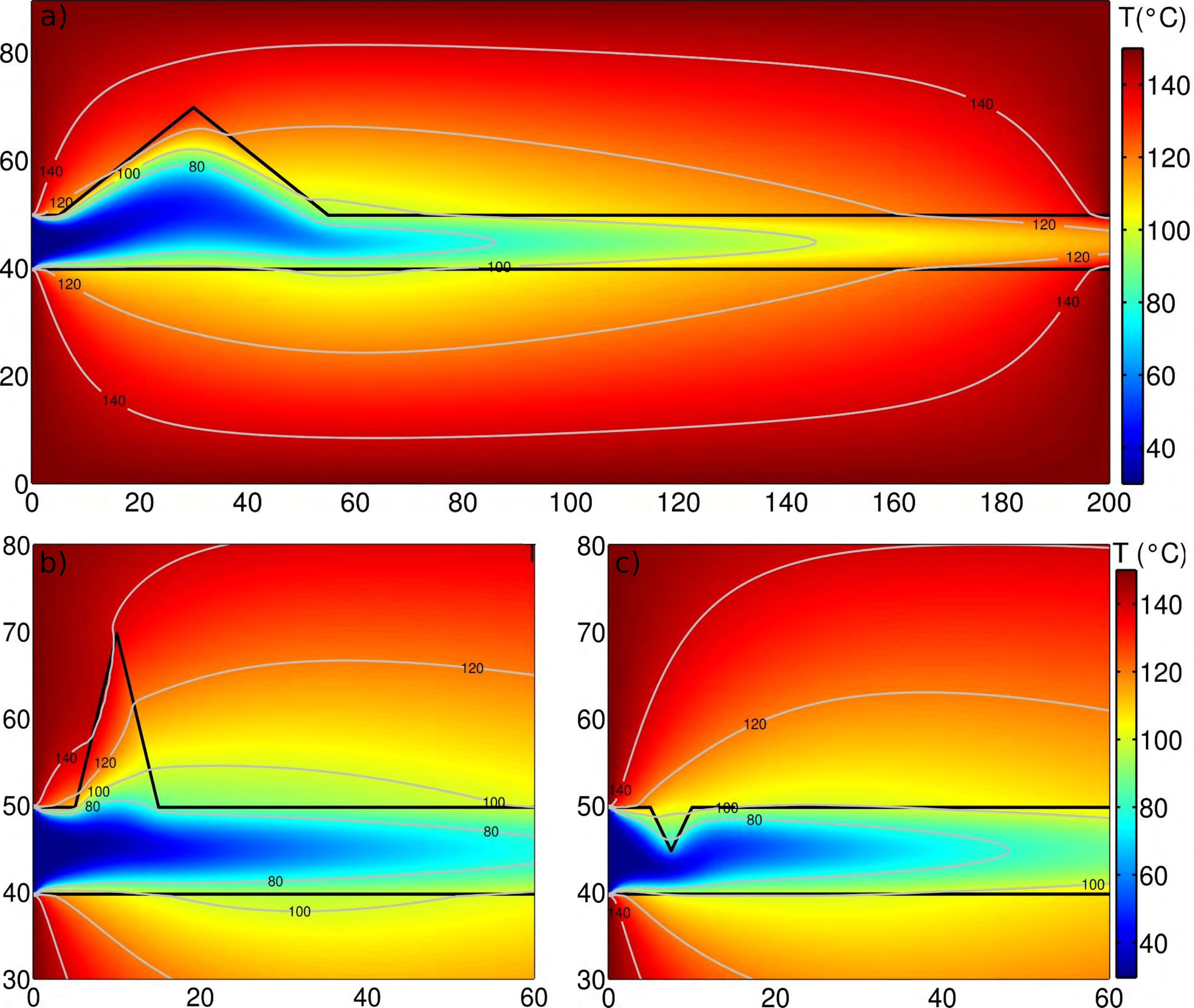}
\caption{Temperature maps at steady state for various asperity shapes:
  (a) $(d, L)=(20\text{~mm}, 50\text{~mm})$, (b) $(d, L)=(20\text{~mm}, 10\text{~mm})$,
  (c) $(d, L)=(-5\text{~mm}, 5\text{~mm})$.  The black lines delimits the
  fracture embedded in the rock. The gray lines are isotherm lines
  (80, 100, 120 and 140\textdegree{}C).\label{fig:TemperatureMap}}
\end{center}
\end{figure}

Figure~\ref{fig:TemperatureMap} shows the temperature map at steady
state corresponding to the hydraulic behavior illustrated in
Fig.~\ref{fig:velocity_h20_L50}.  The shape of the isotherm lines
within the fluid are strongly correlated with the hydraulic flow.  For
$(d, L)=(20\text{~mm}, 50\text{~mm})$, the cold fluid is clearly advected into
the asperity. On the contrary, for $(d, L)=(20\text{~mm}, 10\text{~mm})$, the
deepest part of the asperity is not affected by the injection of the
cold fluid (very low velocities in this separated flow zone), and it
 heats up fast by conduction. As a consequence, the fluid within this
second channel is warmer (ex. compare isolines
$T=80\text{\textdegree{}C}$ on Figs~\ref{fig:TemperatureMap}a-b. For
the narrower fracture, Fig.~\ref{fig:TemperatureMap}c, with $(d,
L)=(-5\text{~mm}, 5\text{~mm})$, the reduction of the hydraulic flow
($H_\text{L\!B}^* = 0.96$ and $H_\text{L\!B}^* = 1.06$ respectively
for $(d, L)=(-5\text{~mm}, 5\text{~mm})$ and $(d, L)=(20\text{~mm}, 50\text{~mm})$)
inhibits the propagation of the cold fluid. This fluid consequently heats
up in a shorter distance (better conductive transport) than for the
two other cases. The rock cools down by conduction: the wall
temperature is inferior to the rock temperature at the border of the
system (150\textdegree{}C), especially where the rock is surrounded by
cold fluid, for instance in Fig.~\ref{fig:TemperatureMap}c, at $(x,
z)=(10\text{~mm}, 45\text{~mm})$.

\begin{figure}[thbp!]
 \begin{center}
 \noindent\includegraphics[width=8.5cm]{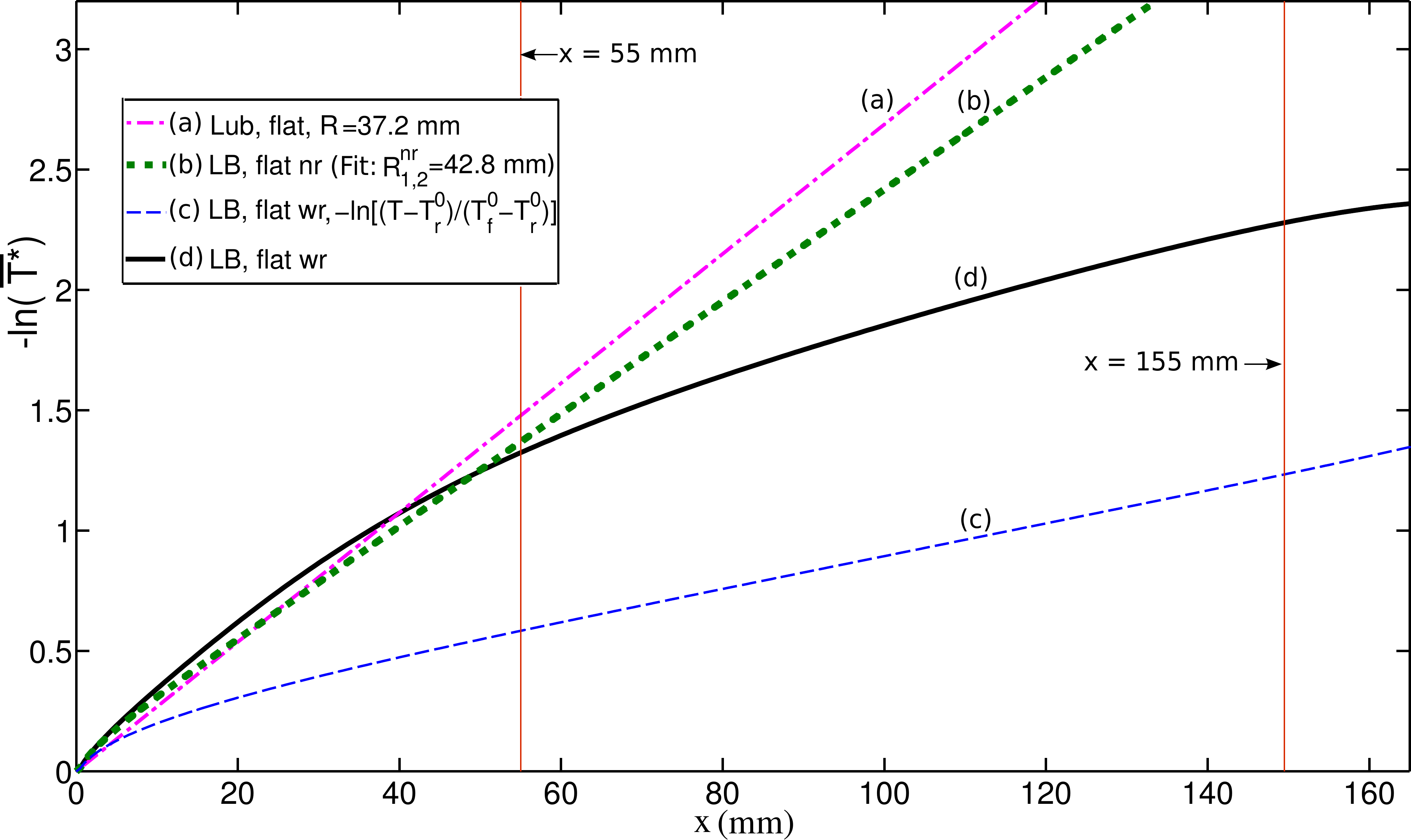}
 \caption{\cvv{(Color online) Plots (a-d) show the opposite of the
   logarithm of the averaged temperature
   $-\ln\left(\overline{T}^{*}\right)$ as a function of $x$ computed
   in a fracture with flat parallel walls separated by $A_0$. The
   temperature is computed (a) with the lubrication approximation, and
   (b, c, d) with the full resolution in LB. ``flat nr'' (b) and
   ``flat wr'' (c-d) stand for computation done within a flat
   geometry, respectively without and with rock temperature
   variation. The vertical lines shows the limits used for the fits
   (whose slopes are respectively $1/R_1$ and $1/R_2$) done on range
   $x\leq 55\text{~mm}$ and $\leq 55 \leq 155 \text{~mm}$.  Plot (c)
   is obtained using the same simulation as plot (d); it differs from
   plot (d) by the way of computing $\overline{T}^{*}$: (c) is
   obtained with the first definition ($T_r^0$ used as reference --
   see text) and
   (d) with the second one (where the variable wall temperature $T_r$
   intervenes).}}
   \label{fig:Tbar_ter}
 \end{center}
 \end{figure}

The thermal behavior is \cvb{now quantified} in the same way as done
in \citet{Neuville10PRE, Neuville10JGI}, i.e. by computing
$\overline{T}^*$ (Fig.~\ref{fig:Tbar_ter}) as in
Eq.~(\ref{eq:Definition:T_2D}).  \cvb{For reference, we first look at
  the results obtained in a fracture with flat parallel
  walls. Fig.~\ref{fig:Tbar_ter}a shows $\ln\overline{T}^*$ obtained
  with lubrication approximation: it is a straight line with a slope
  $1/R_{\text{l\!u\!b}}^{/\!/}$, with
  $R_{\text{l\!u\!b}}^{/\!/}=37.2$~mm. This result is compared to a LB
  simulation performed with an imposed constant temperature rock --
  i.e. only the fluid temperature is computed. This plot
  (Fig.~\ref{fig:Tbar_ter}b) can be very well approximated with a
  single straight line of slope $1/R_{1,2}^{\text{nr}}$, where
  $R_{1,2}^{\text{nr}}=42.8$~mm. This value is higher than
  $R_{\text{l\!u\!b}}^{/\!/}$, i.e. the thermal exchange is worse than
  expected from the lubrication approximation. In the lubrication
  approximation, the in plane diffusion in the fluid is neglected. The
  difference between $R_{\text{l\!u\!b}}^{/\!/}$ and
  $R_{1,2}^{\text{nr}}$ means that the in plane-diffusion tends to
  inhibit the heat exchange.}

\cvb{Then, we relax the hypothesis of constant rock temperature and
  look at the effect of the heat diffusion in the rock: the LB solving
  is done both in the rock and fluid. Two definitions of
  $\overline{T}^*$ are proposed. The first way is to compute it as
  $\left(\overline{T}-T_r^0\right)/\left(T_f^0-T_r^0\right)$
  (Fig.~\ref{fig:Tbar_ter}c). In this expression, the cooling down of
  the rock intervenes only indirectly through its influence on
  $\overline{T}$. This plot is close to a linear plot with slope
  $1/R_1^{0}$ where $R_{1}^{0}=106.2$~mm (fit performed for $x\leq
  55$~mm). This value is much higher than $R_{1,2}^{\text{nr}}$ and
  $R_{\text{l\!u\!b}}^{/\!/}$: it shows that the temperature evolution
  of the rock reduces by more than a factor two the heat
  efficiency. The second way of defining $\overline{T}^*$ takes into
  account the variability of the bottom wall temperature, $T_r$, which
  is computed as $T_r(x)=T(x,z=39.5\text{~mm})$. This definition of
  $\overline{T}^*$ (Fig.~\ref{fig:Tbar_ter}d) emphases the dynamic of
  the fluid temperature compared to the wall temperature. Contrary to
  plots a-c, plot d of Fig.~\ref{fig:Tbar_ter} is not a straight
  line. The beginning of this plot (for $x\leq55$~mm) is approximated
  by a linear fit of slope $1/R_1^{\text{wr}}$, where
  $R_1^{\text{wr}}=42.4$~mm.  The concave curvature of plot (d) of
  Fig.~\ref{fig:Tbar_ter} around $x=55$~mm however attests a change of
  thermal regime. The second part of the curve is fitted with a
  straight line of slope $1/R_2^{\text{wr}}$, where
  $R_2^{\text{wr}}=100.2$~mm.  By choosing the Péclet number equal to
  45.96, there is a good agreement between the thermal lengths
  $R_{\text{l\!u\!b}}^{/\!/}$ and $R_1^{\text{wr}}$. Doing so, we have
  a reference case where the lubrication assumptions holds for the
  computation of the average temperature, in the vicinity of the inlet
  of the fracture.  The P\'eclet number significantly influences the
  agreement between $R_{\text{l\!u\!b}}^{/\!/}$ and $R_1^{\text{wr}}$
  values. At higher values (e.g. $\text{P\!e}=500$) the thermal
  lubrication approximation clearly loose its validity. Further in the
  fracture, the thermal length $R_2$ is higher than $R_1^{\text{wr}}$:
  the thermal exchange efficiency between the wall and the fluid is
  lower than around the injection zone.  This change of regime is not
  predicted by the lubrication approximation, and does not appear with
  an imposed wall temperature. Here, the rock temperature $T_r$
  evolves over time, and is not anymore uniform at stationary regime
  along the fracture: this spatial variability leads to a spatial
  change of regime in the fluid temperature.}

\begin{figure}[thbp!]
\begin{center}
\noindent\includegraphics[width=8.5cm]{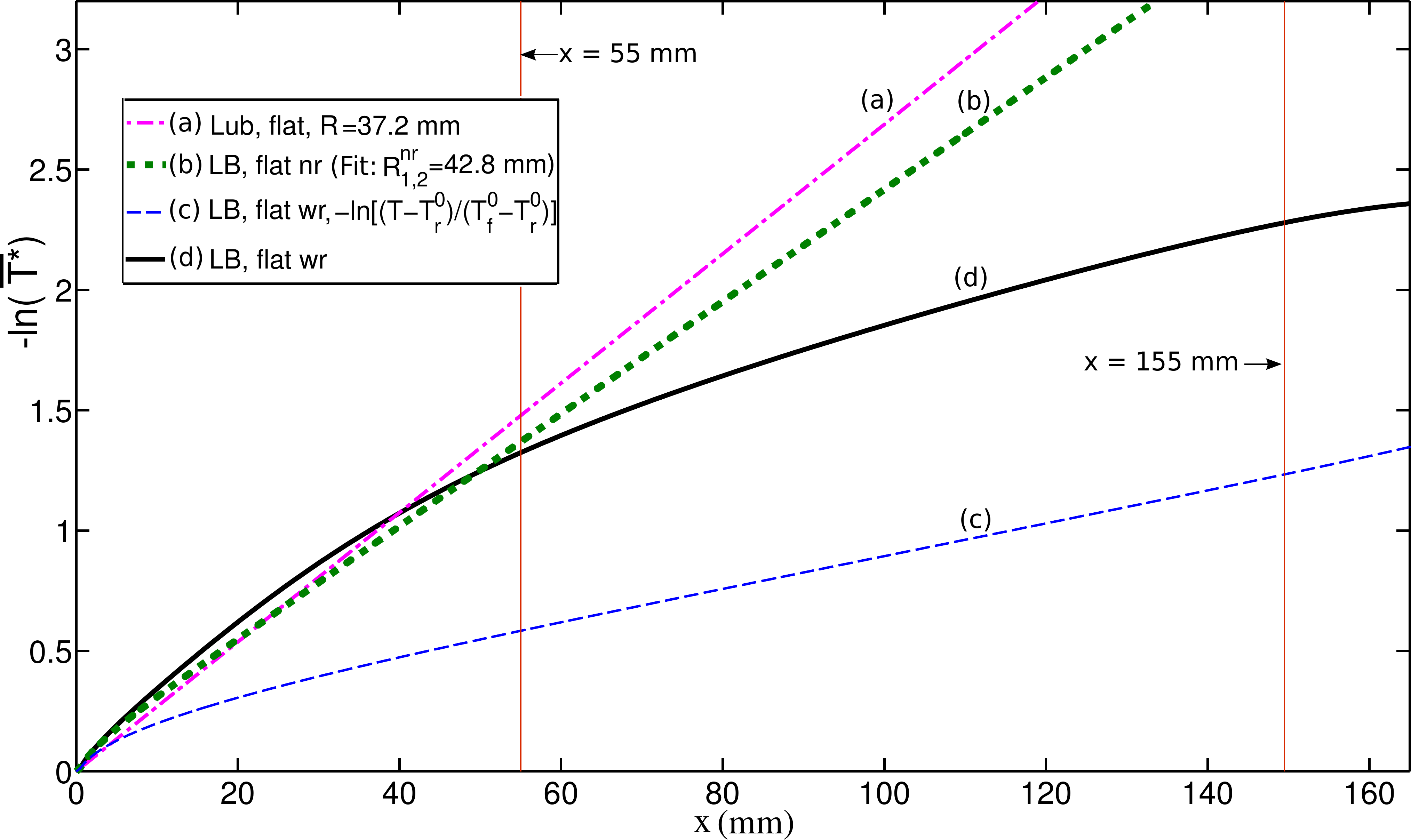}
\caption{ \label{fig:Tbar_ter_b}(Color online) Opposite of the
  logarithm of the averaged temperature
  $-\ln\left(\overline{T}^{*}\right)$ as a function of $x$ computed
  with the full resolution (a, b, c) or with lubrication approximation
  (a', b', c'), for various aperture geometries ($d$ and $L$ indicated
  in the caption). Plot (b'') is obtained with the same equation as
  (b') (Eq.~(\ref{eq:Sol_Tbar_explicit})), with the same $(d,L)$
  values, but with a different $R$ value.  Plots (d) and (e) are the
  linear fits of (a-c, a'-c') over $x \leq 55$~mm. \cvb{The vertical
    lines shows the limits used for the fits.}}
\end{center}
\end{figure}

\cvb{Hereafter, the average temperature $\overline{T}^*$ in LB
  simulations is computed with the second definition (i.e. using the
  space variable wall temperature $T_r(x)=T(x,z=39.5\text{~mm})$), for
  other geometries (Fig.~\ref{fig:Tbar_ter_b}). The full resolution
  solving (plots a, b, c) are compared to the lubrication
  approximation (a', b', c').} In general, the lubrication
approximation gives very different results, especially for large
$x$. For negative $d$ values \cvbb{(bumps)}, the LB computation shows
that the temperature behavior changes at the abscissa corresponding to
the edge of the corner ($x_0+L/2$).  Within the lubrication, it is
possible, by adapting $R$ in Eq.~(\ref{eq:Sol_Tbar_explicit}), to
obtain a similar behavior for $x<x_0+L/2$, but the change of slope in
$x=x_0+L/2$ cannot be modeled in this way, \cvv{as attested by the
  poor quality of the fit shown in plot (d) of
  Fig.~\ref{fig:Tbar_ter_b}.}  This change \cvbb{of slope }might be
linked to the change of the hydraulic flow, also occurring around the
corner edge, and not predicted by the lubrication approximation.  For
positive $d$ values \cvbb{(hollow asperity)}, the corner geometry
causes smoother variations in the slope of $-\ln(\overline{T}^*)$,
than expected with the lubrication approximation. The incomplete
modeling of the heat diffusion artificially implies sharper
temperature variations. Some similarities in the variations can
however been observed, notably for the highest values of $(d,L)$ (e.g.
Fig.~\ref{fig:Tbar_ter_b}b, b', b''), close to the injection
zone ($x\leq 55$~mm), providing that the thermal length is adjusted
(32.3~mm instead of 46.2~mm for $(d,L)=(20\text{~mm}, 50\text{~mm})$).

The quantification of the thermal behavior is done in the same way as
previously (cf \ref{sec:therm-lubr-appr}). Two thermal lengths $R_1$
and $R_2$ are defined, by approximating $-\ln(\overline{T}^*)$ with
two linear fits on two $x$ ranges.\label{def_R1} $R_1$ and $R_2$
(reported in Tab.~\ref{tab:Lref}) respectively characterizes the
thermal behavior in the vicinity, and far from the injection zone,
(i.e.  for $x \leq 55$~mm and $55\leq x \leq 150$~mm). These values
are commented in Sec.~\ref{sec:results:-expl-param}. \cvv{The limit of
  55~mm corresponds to the change of behavior observed with the flat
  fracture in LB (Fig.~\ref{fig:Tbar_ter}d).  The range $x \leq 55$~mm
  also systematically includes the asperity: the temperature estimate
  in $x=55$~mm somehow reflects what would be observed by a
  temperature probe located there, investigating at the integrated
  effect of the morphology.}

\begin{center}
\begin{table}
\small
 \begin{tabular}{c|c|c|c|c|c|c|c|c}
   \hline
   &\multicolumn{2}{c|}{lubrication}&\multicolumn{2}{c|}{LB}&\multicolumn{4}{c}{ratios}\\
   \hline
   $d,L$ & $R$ &$R_{1\text{l\!u\!b}}$ & $R_1$ &
   $R_2$& { $\frac{R_1}{R_{\text{l\!u\!b}}^{/\!/}}$} &
   { $\frac{R_1}{R_{1\text{l\!u\!b}}} $} & {$\frac{R_2}{R_{\text{l\!u\!b}}^{/\!/}}$} &
    {$\frac{R_2}{R} $}\\
   \hline
   flat wr &37.2 &37.2& 42.4 &100.2&1.14
   &1.14&2.70&2.69 \\
   flat nr &37.2 &37.2& 42.8 &42.8 &1.15
   &1.15&1.15&1.15 \\
   \hline 
   20, 10 & 38.7& 41.7 &45.5& 98.0&1.22&
   1.09&2.63&2.53\\
   \hline
   20, 50 &46.2 &92.3 &64.8 & 101.5 & 1.74 & 0.70 & 2.72&2.20\\
   \hline
   -5, 5 & 35.4& 34.6& 38.5& 91.7&1.04&
   1.11&2.46 &2.59\\
   \hline
   -5, 50 &24.8 &17.0 &26.6 &79.3 & 0.72 &
   1.58 & 2.1 &3.20\\
   \hline
\end{tabular}
\caption{Thermal lengths obtained for various triangular geometries,
obtained from Eq.~(\ref{eq:LengthThermalizationParallelPlates}) (for
$R$) and from the fits of the curves $\ln\overline{T}^*$ (cf
Fig.~\ref{fig:Tbar_ter} and Fig.~\ref{fig:Tbar_ter_b}).  Subscripts 1
and 2 refer to the range of values where the fit has been done ($
x\leq 55$~mm and $55\leq x\leq 150$~mm). ``flat wr'' and ``flat nr''
stand for flat geometry with and without rock temperature variation}
\label{tab:Lref}
\end{table}
 \end{center}

\normalsize

\begin{figure}[thbp!]
\begin{center}
\noindent\includegraphics[width=8.cm]{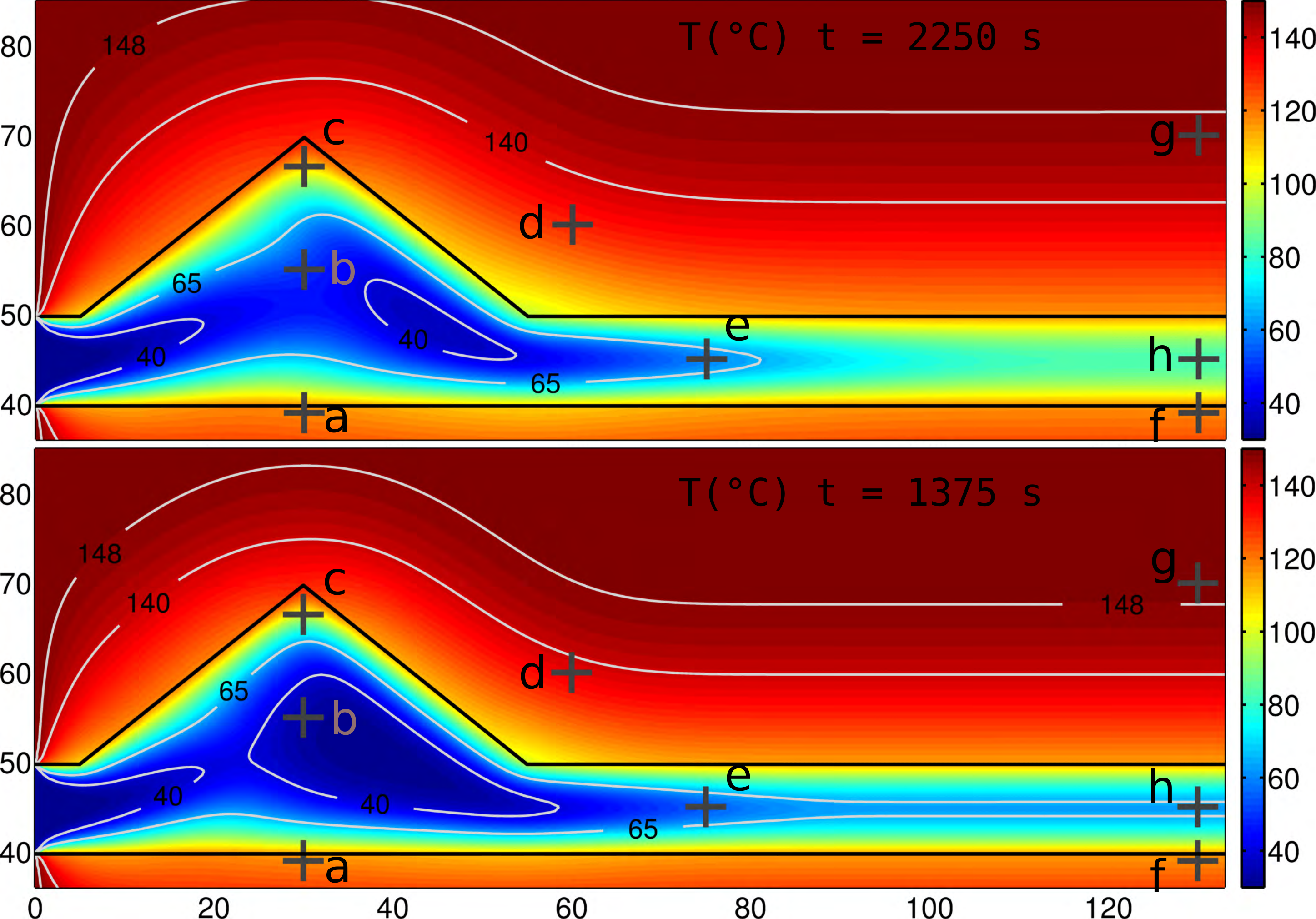}
  \caption{(Color online) Temperature as a function of time, for the
geometry characterized by $(d,L)=(20\text{~mm}, 50\text{~mm})$, at $t=1375$~s and
$t=2250$~s, with isotherm lines (40, 65, 140, 148)\textdegree{}C. The
letters a-h indicates the locations where the temperature evolution is
observed in Fig.~\ref{fig:temperature_time}.
    \label{fig:temperature_time_aplat}}
\end{center}
\end{figure}

\begin{figure*}[thbp!]
\begin{center}
\noindent\includegraphics[width=9.cm]{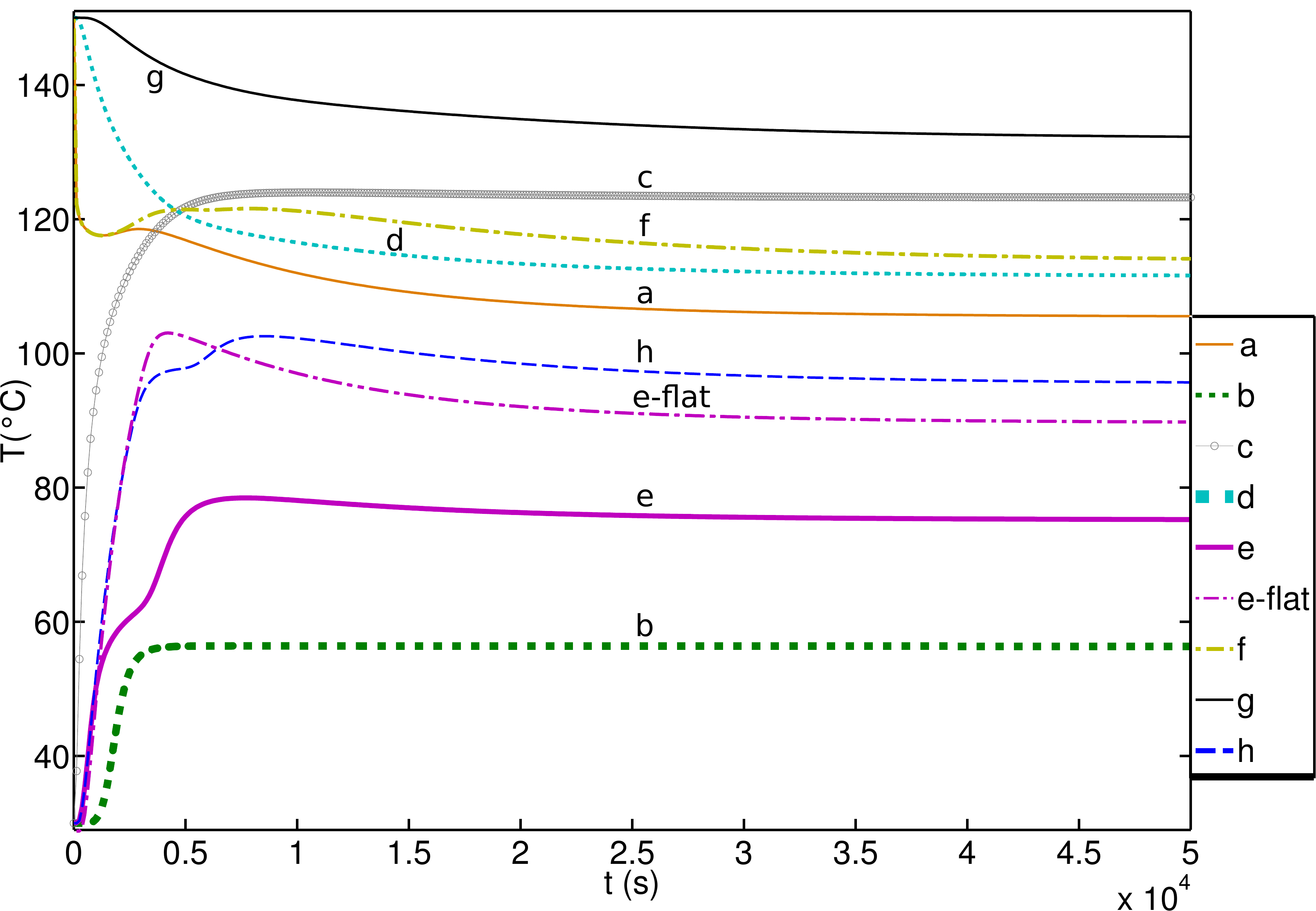}
  \caption{(Color online) Temperature as a function of
    time, at different locations a-h,  for the geometry
   characterized by $(d,L)=(20\text{~mm}, 50\text{~mm})$, and for the flat geometry
   (caption: \cvbb{e-flat}). The locations, indicated by the letters
    a-h, are shown in Fig.~\ref{fig:temperature_time_aplat}.
    \label{fig:temperature_time}}
\end{center}
\end{figure*}

The temperature of the fluid and rock evolves over time
(Fig.~\ref{fig:temperature_time_aplat}).  Because the diffusivity of
the fluid is much higher than that of the rock, first, the fluid warms
up fast (\cvbb{points and plots b, c, e, h in}
Fig.~\ref{fig:temperature_time}). On the contrary, the points in the
rock which are close to the fluid, cool down fast (plots a, f of
Fig.~\ref{fig:temperature_time}). With a slower dynamic (intermediate
time scale) the heat source maintaining the system borders at a hot
temperature, provides a heat flux that diffuses towards the walls
(points a and f), which causes their temperature to increase
again. The temperature at points located far enough from the fluid (d,
g), does not change as fast at early times; points (d, g) simply cool
down in a monotonic way. Similar variations are observed for a flat
fracture.  The sudden slow down of the heating process of points e and
h (around 2000 and 4000~s), located after the asperity can however
specifically be attributed to the asperity (see for comparison, plot
\cvbb{e-flat, obtained in a flat geometry}). This variation is not
observed at points (b) and (c), located in the fluid, close to the
injection zone. Finally the temperature field decreases everywhere at
very low time scale, and the system seems to reach a steady state. The
variation of the temperature field over time is complex as, two points
close to each other, may have different variations. Some points reheat
and cool down alternately several times, which makes it difficult to
forecast the extracted heat.


\section{Results: exploration of the parameter space}
\label{sec:results:-expl-param}
\subsection{Hydraulic aperture and thermal length computation}
\begin{figure*}[thbp!]
\begin{center}
\noindent\includegraphics[width=\linewidth]{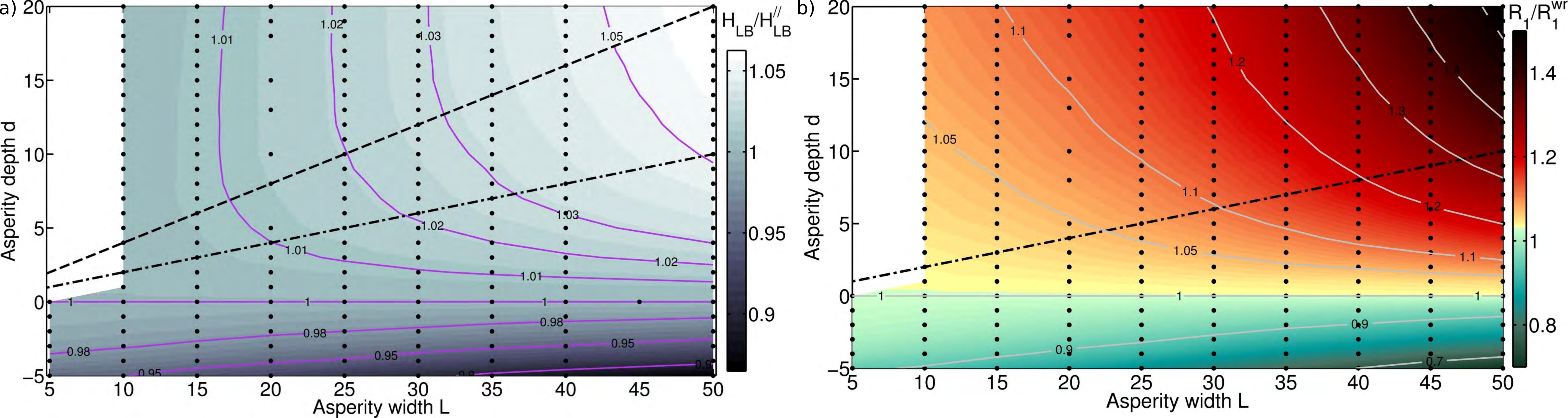}
\caption{(Color online) Color maps of the normalized hydraulic
  aperture $H_{\text{L\!B}}^*$ (a) and normalized thermal length
  $R_1/R_1^{\text{wr}}$ (b) as a function of the asperity depth
  $d$ and width $L$.  $H_{\text{L\!B}}^*$, and $R_1$, $R_1^{\text{wr}}$ are respectively defined in
    paragraphs~\ref{def_HLBetoile} and \ref{def_R1}. The
    normalizations are done by constant values. The dots indicate the
    parameters for which the values are computed. The gray lines are
    isolines.  The black dashed and dot-dashed lines respectively are
    the lines $d=0.4 L$ and $d=0.2 L$.
\label{fig:Stat_hydrau1}\label{fig:StatThermal1_normCst}}
\end{center}
\end{figure*}

Exploration of the parameter space for $d$ and $L$, and their
influence on the hydraulic aperture and the thermal length has been
investigated. Figure~\ref{fig:Stat_hydrau1} shows a color map of the
normalized hydraulic aperture $H_{\text{L\!B}}^*$ as a function of $d$
and $L$. For $d>0$ (channel larger than $A_0$\cvbb{with a hollow asperity}), the permeability
increases compared to a flat channel of aperture $A_0$, as
$H_{\text{L\!B}}>H_{\text{L\!B}}^{/\!/}$. For $d<0$ (channel
narrower than $A_0$ \cvbb{with a bump}), the opposite behavior is observed
$H_{\text{L\!B}}<H_{\text{L\!B}}^{/\!/}$. \cvv{It is possible to
  divide the map in three areas that can be approximately separated
  with two straight lines.}  \pv{This computation also shows that
  f}For $d>0.4 L$ (black dashed line), the hydraulic aperture for a
given width $L$ tends to be constant (vertical isolines on
Fig.~\ref{fig:Stat_hydrau1}) whatever the depth $d$ of the asperity
is. On the contrary, for $d<0.2 L$ (black dash-dotted line), the
hydraulic aperture tends to be constant for a given $d$ (horizontal
isolines). Among the explored $\beta$ angles (defined as $\arctan
(2d/L)$, in the range $0<\beta \leq 360$\textdegree{}, not
exhaustively explored), these limits corresponds to the angles \label{Mofat_straight_lines}
\pv{$\beta<102.7$\textdegree{}} \cvv{$\beta<103$\textdegree{}} ($d>0.4
L$) and $\beta>136$\textdegree{} ($d<0.2 L$). This last limit angle is of same order as the one obtained by
\citet{Moffatt64}\cvv{, whose study was done using slightly different
  flow assumptions.  He showed that even at vanishing Reynolds number, eddies form in a corner between two intersecting planes when the angle exceeds 146\textdegree{}, when a shear flow is imposed far away from the corner. The presence of eddies is well supported in
  our case by almost zero velocity values and/or negative velocities
  observed in the middle of the corner (see
  Fig.~\ref{fig:velocity_h20_L10}e). The flow in the corner is very
  small, almost separated from the main flow; it thus does not
  contribute significantly to the hydraulic flow.}

Figure~\ref{fig:StatThermal1_normCst} shows a map of the thermal
lengths $R_1$ (obtained from LB computation for range $x \leq 55$~mm),
normalized by $R_1^{\text{wr}}$ (constant value) as a function of $d$
and $L$ values. For any geometry with $d>0$ (hollow asperities), the
thermal exchange around the asperity is inhibited compared to that
within a flat fracture ($R_1>R_1^{\text{wr}}$). For $d<0$, the thermal
exchange in on the contrary better ($R_1<R_1^{\text{wr}}$). For
geometry with angle $\beta>136$\textdegree{} ($d<0.2 L$), for a given
depth $d$, $R_1$ shows few variations.  For the range $55\leq x \leq
150$~mm, the thermal length $R_2$ was also computed. $R_2$ is on
average 2.3 times higher than $R_1^{\text{L\!B}}$: the thermal
exchange is far less efficient than it is close to the injection
zone. \cvbb{Far from the injection zone, the thermal exchange} also does not vary much with the geometry. Indeed, on
average, $R_2/R_2^{\text{wr}}=0.98$, with a standard deviation of
0.04).

\subsection{Comparison to the lubrication approximation}

It is questionable if our parameter study may be compared to the
results obtained in \citet{Neuville10PRE}.  This latest study focused
on the hydraulic aperture and thermal length obtained for self-affine
fractures under lubrication approximation, using the analysis exposed
in \ref{sec:with-lubr-appr}. By contrast with the current study, the
aperture studied in \citet{Neuville10PRE} is self-affine, which means
that using the Fourier decomposition, it is decomposed in
$a(x,y)=\sum_k \tilde{a}(k_x,k_y) e^{-2i\pi (k_x x + k_y y)}$, where
$\boldsymbol{k}$ is the wave vector and $\tilde{a} (k_x,k_y)$ scales
as $\tilde{a}(k_x,k_y) \sim C k^{-1-\zeta}$ for $k\neq 0$. For such
aperture, it was shown in \citet{Neuville10JGI} that the hydraulic and
thermal behavior can mostly be deduced from the highest wave lengths
of the aperture. For a flat aperture perturbed with an isolated
triangular shape, the situation is very different. Its power spectrum
at the largest length scales not only depends on the triangular
asperity shape, but also depends on the length of the flat area before
and after the asperity.  For a given pressure gradient, performing
statistics (like calculating the hydraulic aperture, mean aperture,
and standard deviation) at the asperity scale or over the full
fracture scale provides very different results.  We clearly see that
$H_{\text{L\!B}}/A$ is dominated by the range where it is calculated,
and are therefore difficult to be compared with the values obtained
under lubrication in \citet{Neuville10PRE}.  For this reason, we
compared (Fig.~\ref{fig:Stat_hydrau1}) the hydraulic aperture
$H_{\text{L\!B}}^*$ to the one obtained under lubrication
approximation $H_{\text{l\!u\!b}}^*$, defined as done in
\ref{def_ratio_H_Hlubri}.

\begin{figure}[bhp!]
\begin{center}
\noindent\includegraphics[width=\linewidth]{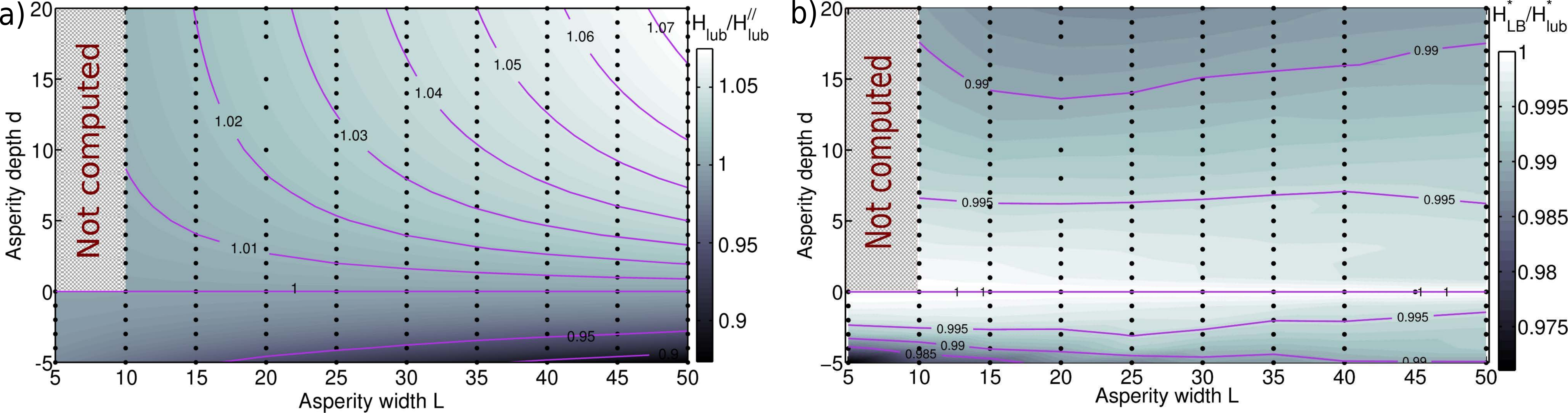}
\caption{\cvv{(Color online) Color maps as a function of the asperity
    depth $d$ and width $L$ of (a) the hydraulic aperture computed
    with the lubrication approximation,
    $H_{\text{l\!u\!b}}^*=H_{\text{l\!u\!b}}/H^{/\!/}_{\text{l\!u\!b}}$,
    and (b) the ratio between the hydraulic aperture computed with
    LBM, $H_{\text{L\!B}}^*$, and the one computed with the
    lubrication approximation, $H_{\text{l\!u\!b}}^*$ . The dots
    indicate the computed values from which the
    map is deduced. The purple lines are isolines.
\label{fig:Stat_hydrau_lubri}\label{fig:Stat_hydrau2}}}
\end{center}
\end{figure}

\cvv{Figure~\ref{fig:Stat_hydrau_lubri}a shows the hydraulic aperture under
  lubrication $H^*_{\text{l\!u\!b}}$. It can be noticed that the
  isoline shapes differ from that shown in
  Fig.~\ref{fig:Stat_hydrau1}: the hydraulic aperture computed under
  lubrication approximation evolves smoothly for a constant depth $d$
  or width $L$. It is also not possible to delimit the different zones
  separated by the straight lines as done for the LBM computation in
  paragraph \ref{Mofat_straight_lines}.} Figure \ref{fig:Stat_hydrau2}b
shows $H^*_{\text{L\!B}}/H^*_{\text{l\!u\!b}}$. This ratio is always
very close to 1, with a systematic characteristic:
$H^*_{\text{L\!B}}<H^*_{\text{l\!u\!b}}$. It means that the hydraulic
flow is very slightly overestimated with the lubrication
approximation. However, it should be kept in mind, that the hydraulic
aperture only reflects an averaged permeability, and not the local
flow differences. We will now see if these local differences may be
seen through the thermal behavior.

\begin{figure*}[thbp!]
\begin{center}
\noindent\includegraphics[width=\linewidth]{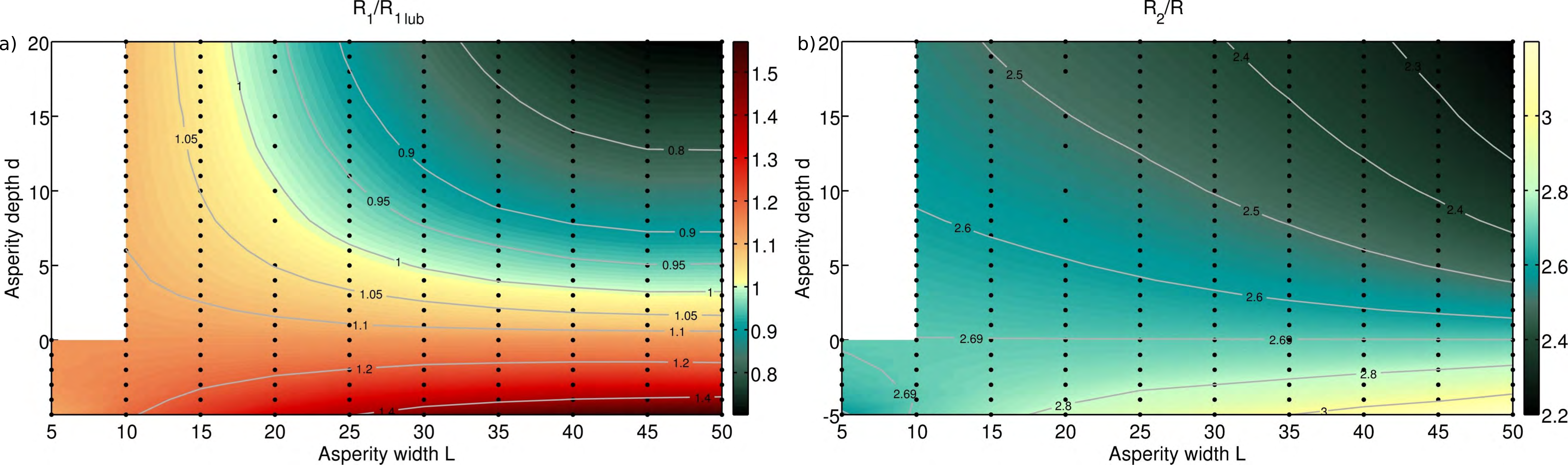}
\caption{(Color online) Color maps of (a) the normalized thermal
  length $R_1/R_{1\,\text{l\!u\!b}}$ ($R_1$ computed for $x \leq
  55$~mm) and (b) $R_2/R$ (computed for $ 55 \leq x \leq
  150\text{~mm}$) as a function of the asperity depth $d$ and width
  $L$. $R_{1,2}$, $R_{1 \text{l\!u\!b}}$ and $R$ are respectively
  defined in paragraphs~\ref{def_R1}, \ref{def_R1lub} and in
  Eq.~\ref{eq:LengthThermalizationParallelPlates}. The dots indicate
  the computed values from which the map is deduced.  Isolines for
  $R_{1\,\text{or}\,2}^*$ are
  shown. \label{fig:StatThermal_norm_lubri_variable}}
\end{center}
\end{figure*}
Figure~\ref{fig:StatThermal_norm_lubri_variable} shows a map of the
thermal lengths $R_1$, normalized by $R_{1\,{\text{l\!u\!b}}}$, which
also varies with the geometry, as a function of $d$ and $L$
values. The domain can be separated in two. First the geometries where
$R_1/R_{1\,{\text{l\!u\!b}}}<1$: for these geometries, the thermal
exchange is actually better than expected from the lubrication
approximation. These geometries correspond to asperities which are
deep and large enough. This can be explained by the fact that the
lubrication approximation does not take into account the local
reduction of the hydraulic flow within the corner, which diminishes
the convective heat transport, and therefore favorites a better local
heating up, by conduction. Second, the domain where
$R_1/R_{1\,{\text{l\!u\!b}}}>1$: for these geometries the thermal exchange is not as good
as expected from the lubrication approximation. Note that, for
the geometries with $d<0$, $R_1$ is overestimated, as it is obtained from
a fit done for $x\leq 55$~mm, while a change of regime occurs in the
middle of the asperity (the thermal exchange is observed to be less
efficient for $x\geq x_0+L/2$).

\section{Discussion and conclusion}

Our hydraulic results are coherent with literature: quasi stagnant
fluid is observed in the corners with small $\beta$ angles. For the
studied corner geometry, the hydraulic aperture obtained with the full
resolution, although not very different, is systematically smaller
than the one obtained with lubrication assumptions. Compared to
  fractures with parallel flat walls, the hydraulic aperture is
  systematically higher for the fractures with the triangular hollows,
  and smaller for bump asperities: this is completely expected as the
  aperture is clearly either increased or reduced. For the asperities
  with $\beta$ angle superior to 136\textdegree{}, the hydraulic
  aperture depends almost only on the depth of the asperities, while
  for $\beta$ angle inferior to 103\textdegree{}, the hydraulic
  aperture depends almost only on the width of the asperities. The
  fluid trapped in deep asperities does not mix easily with the main
  flow. In order to mix better this trapped fluid, and prevent
  stagnant fluid, it could be interesting to stimulate the system with
  oscillating pressure gradients. The oscillating frequency should be
  smaller than the diffusion time in the corner, i.e. smaller than
  $d^2/\nu$.

  For the heat exchange, we notice that the deep part of the corner,
  although insignificant in term of hydraulic flux contribution,
  changes the heat exchange. Compared to fractures with parallel flat
  walls, the heat efficiency is systematically worse (factor 1 to 1.4)
  for the studied fractures with hollow asperities, and better for the
  bumpy ones (factor 0.7 to 1). The hydraulic aperture only depend on
  the width of the asperity for $\beta<103$\textdegree{}. This feature
  is not recovered for the thermal lengths: this shows that the
  thermal exchange is even more dependent on the geometry shape than
  the hydraulic behavior. The fluid trapped in the corner is mostly
  sensitive to heat diffusive processes whose time scale depend of
  $\sqrt (d_f^2/\chi_f)$ and $\sqrt (d_i^2/\chi_f)$, where $d_f$ is
  the distance to the well flowing fluid and $d_i$ is the distance to
  the fluid-rock interface. It means that the deepest point in the
  corner, which is in contact with the warm rock, transmits the heat
  of the rock to the main flowing fluid at latest after $\Delta
  t=d^2/\chi_f$, where $d$ is the depth of the corner. When the main
  flow penetrates deeper in the corner, the distance between the
  flowing fluid and warm wall is reduced: we may think that this
  transfer time is reduced. This is however not necessary true since
  the volume of fluid trapped in the corner, as well as the surface of
  exchange (interface fluid rock) also change and modify the heat
  exchange. For instance very deep and thin asperities transmit very
  efficiently the heat from the rock. It is important to note that the
  temperature efficiency can however not be directly linked neither to
  the volume of the asperity nor to the ratio of the volume and
  surface of exchange.  Another important parameter is the distance
  $d_s$ between the heat source and the fluid-rock interface: the
  variations of the wall temperature are sensitive to both time scales
  $\sqrt(d_f^2/\chi_f)$ and $\sqrt (d_s^2/\chi_r)$. If both time
  scales are of same order, the temperature fluctuates quickly and is
  very sensitive to local heterogeneities. On the field, it is very
  likely that $d_s>>d_f$: the diffusion process transports the heat
  through long distances in the rock. This will bring slower dynamic
  in the temperature variation. Local temperature heterogeneities due
  to complex geometry of fracture will however still behave as local
  sources (cooler or warmer sites than the surroundings) and create
  short time scale variations.  The forecast of pumped water
  temperature should therefore not relies on simple parameters, but
  should really take into account the geometry of the porous
  medium. The diffusive exchange is important not only in the corner,
  but also more generally, close to the fluid rock interface, where
  the velocity is low. The advective heat transfer mostly occurs in
  the middle of the channel, where the velocity is high; this transfer
  is characterized by time scales of order of $L_x/u$. In order to
  better mix the fluid between the zones where the advective process
  is efficient and those where the diffusive process occurs, it would
  be interesting to introduce some tortuosity in the fracture (with a
  typical length scale of order $\chi_f/u$), or to stimulate the
  system by oscillating the pressure gradient, with a time scale
  smaller than $\sqrt(d/\chi_f)$. The oscillations may locally
  introduce changes of direction of the flow, and transverse velocity
  components, which may very efficient mix the fluid. The computed
  thermal lengths can be associated to heat efficiencies. The heat
  exchange efficiency may also be defined in other ways, by computing
  the difference of the total energy flux between the inlet and outlet
  of the fluid.

  The full resolution of the temperature field computed with the
  lattice Boltzmann method shows that the heat efficiency evolves with
  the distance to the inlet of the fracture. This evolution can be
  attributed both to the asperity, and the cooling of the rock. Two
  thermal lengths were therefore defined to evaluate the heat
  efficiency. Far enough from the inlet, the thermal lengths obtained
  with lubrication approximation are clearly underestimated (of a
  factor 2.2 to 3.2), which means that the heat efficiency is
  overestimated. Close to the injection point, the efficiency is
  either under or over estimated, depending on the shape of the
  asperity. For deep and thin asperities, as well as rather flat
  asperities (asperities with small volume), and bump asperities, the
  thermal exchange efficiency is underestimated with the lubrication
  approximation of a factor 1 to 1.6. It is otherwise slightly
  overestimated (factor 0.7 to 1).

  The time variations of the heating and cooling of the fluid and the
  rock have also been studied. It was observed local and sudden slow
  downs of the heating process, probably caused by the
  asperity.
A similar phenomena with a more complicated geometry could explain the
sudden temperature variations during transient regime observed when
pumping in geothermal systems.  

In spite of the simplicity of the studied fracture geometry, the
observed hydro-thermal exchanges are finally very complex. The local
three-dimensional phenomena modify both the hydraulic and thermal
macroscopic properties, in a different way. To extend this work on
real field, it would be interesting to consider multiplicity of scales
-- either a distribution of asperity sizes, or a network of fractures
--. At the field scale, it is not clear how much time will be
necessary for a real steady state to be reached. This depends on the
volume of the hydraulically stimulated rock, and also on the distance
and time dependency of the heat sources.


%
%
%
%
%
%

%
%
%
%

\begin{acknowledgments}
  We wish to thank K.J. Måløy, M. Erpelding, A. Cochard, F. Renard,
  L. Talon and O. Aursjø for fruitful discussions.  We acknowledge the
  financial support of the Research Council of Norway through the
  YGGDRASIL mobility grant for the project n${^\circ}$202527, the
  PETROMAKS project, the PICS program France-Norway, the ANR
  Landquake, the ITN FLOWTRANS, and the support of the CNRS INSU. We
  also thank the French network of Alsatian laboratories, REALISE.
\end{acknowledgments}

%
%

%
%

\begin{thebibliography}{62}
\providecommand{\natexlab}[1]{#1}
\expandafter\ifx\csname urlstyle\endcsname\relax
  \providecommand{\doi}[1]{doi:\discretionary{}{}{}#1}\else
  \providecommand{\doi}{doi:\discretionary{}{}{}\begingroup
  \urlstyle{rm}\Url}\fi

\bibitem[{\textit{Amaziane et~al.}(2008)\textit{Amaziane, El~Ossmani, and
  Serres}}]{Amaziane08}
Amaziane, B., M.~El~Ossmani, and C.~Serres (2008), Numerical modeling of the
  flow and transport of radionuclides in heterogeneous porous media,
  \textit{Computational Geosciences}, \textit{12}(4), 437--449.

\bibitem[{\textit{Andrade~Jr. et~al.}({2004})\textit{Andrade~Jr., Henrique,
  Almeida, and Costa}}]{Andrade04}
Andrade~Jr., J., E.~Henrique, M.~Almeida, and M.~Costa ({2004}), Heat transport
  through rough channels, \textit{Physica A}, \textit{{339}}({3-4}), 296,
  \doi{{10.1016/j.physa.2004.03.066}}.

\bibitem[{\textit{Bergbauer et~al.}(1998)\textit{Bergbauer, Martel, and
  Hieronymus}}]{Bergbauer98}
Bergbauer, S., S.~Martel, and C.~Hieronymus (1998), Thermal stress evolution in
  cooling pluton environments of different geometries,
  \textit{Geophys.~Res.~Lett.}, \textit{25}(5), 707--710.

\bibitem[{\textit{Bernab\'e and Olson}(2000)}]{Bernabe00}
Bernab\'e, Y., and J.~Olson (2000), The hydraulic conductance of a capillary
  with a sinusoidally varying cross-section, \textit{Geophys.~Res.~Lett.},
  \textit{27}(2), 245--248.

\bibitem[{\textit{Bhatnagar et~al.}(1954)\textit{Bhatnagar, Gross, and
  Krook}}]{Bhatnagar54}
Bhatnagar, P., E.~Gross, and M.~Krook (1954), A model for collision processes
  in gases. {I.} small amplitude processes in charged and neutral one-component
  systems, \textit{Phys.~Rev.}, \textit{93}(3), 511--525.

\bibitem[{\textit{Bouchaud}(1997)}]{Bouchaud97}
Bouchaud, E. (1997), Scaling properties of cracks, \textit{J. Phys.: Condens.
  Matter}, \textit{9}, 4319--4344.

\bibitem[{\textit{Boutt et~al.}(2006)\textit{Boutt, Grasselli, Fredrich, Cook,
  and Williams}}]{Boutt06}
Boutt, D., G.~Grasselli, J.~Fredrich, B.~Cook, and J.~Williams (2006), Trapping
  zones: The effect of fracture roughness on the directional anisotropy of
  fluid flow and colloid transport in a single fracture,
  \textit{Geophys.~Res.~Lett.}, \textit{33}(21).

\bibitem[{\textit{Brown}(1987)}]{Brown87}
Brown, S. (1987), Fluid flow through rock joints: The effect of surface
  roughness, \textit{J.~Geophys.~Res.}, \textit{92}(B2), 1337--1347.

\bibitem[{\textit{Brown and Scholz}(1985)}]{Brown85}
Brown, S., and C.~Scholz (1985), Broad bandwidth study of the topography of
  natural rock surfaces, \textit{J.~Geophys.~Res.}, \textit{90}(B2),
  12,575--12,582.

\bibitem[{\textit{Brown et~al.}(1995)\textit{Brown, Stockman, and
  Reeves}}]{BrownStock95}
Brown, S., H.~Stockman, and S.~Reeves (1995), Applicability of the {R}eynolds
  equation for modeling fluid flow between rough surfaces,
  \textit{Geophys.~Res.~Lett.}, \textit{22}(18), 2537--2540.

\bibitem[{\textit{Brush and Thomson}(2003)}]{Brush03}
Brush, D., and N.~Thomson (2003), Fluid flow in synthetic rough-walled
  fractures: {N}avier-{S}tokes, {S}tokes, and local cubic law simulations,
  \textit{Water Resour. Res.}, \textit{39}(4), 1085.

\bibitem[{\textit{Candela et~al.}(2009)\textit{Candela, Renard, Bouchon,
  Brouste, Marsan, and Schmittbuhl}}]{Candela09}
Candela, T., F.~Renard, M.~Bouchon, A.~Brouste, D.~Marsan, and J.~Schmittbuhl
  (2009), Characterization of fault roughness at various scales: {I}mplications
  of three-dimensional high resolution topography measurements, \textit{Pure
  Appl. Geophys.}, \textit{166}(10--11), 1817--1851.

\bibitem[{\textit{Cardenas et~al.}(2007)\textit{Cardenas, Slottke, Ketcham, and
  {Sharp Jr.}}}]{Cardenas07}
Cardenas, M., D.~Slottke, R.~Ketcham, and J.~{Sharp Jr.} (2007),
  {N}avier-{S}tokes flow and transport simulations using real fractures shows
  heavy tailing due to eddies, \textit{Geophys.~Res.~Lett.}, \textit{34}(14).

\bibitem[{\textit{Chopard and Droz}(1998)}]{Chopard98}
Chopard, B., and M.~Droz (1998), \textit{Cellular Automata Modeling of Physical
  Systems}, University Press, Cambridge.

\bibitem[{\textit{Cvetkovic et~al.}(2004)\textit{Cvetkovic, Painter, Outters,
  and Selroos}}]{Cvetkovic04}
Cvetkovic, V., S.~Painter, N.~Outters, and J.~Selroos (2004), Stochastic
  simulation of radionuclide migration in discretely fractured rock near the
  {{\"{A}sp\"{o}} {H}ard {R}ock {L}aboratory}, \textit{Water Resour. Res.},
  \textit{40}(2), W024,041--W02404,114.

\bibitem[{\textit{d'Humi\`eres et~al.}(1986)\textit{d'Humi\`eres, Lallemand,
  and Frisch}}]{FCHC86}
d'Humi\`eres, D., P.~Lallemand, and U.~Frisch (1986), Lattice gas models for
  3{D} hydrodynamics, \textit{Europhys. Lett.}, \textit{2}(47), 291--297.

\bibitem[{\textit{Drazer and Koplik}(2001)}]{Drazer01}
Drazer, G., and J.~Koplik (2001), Tracer dispersion in two-dimensional rough
  fractures, \textit{Phys.~Rev.~E}, \textit{63}(5 II), 0561,041--05610,411.

\bibitem[{\textit{Drazer and Koplik}(2002)}]{Drazer02}
Drazer, G., and J.~Koplik (2002), Transport in rough self-affine fractures,
  \textit{Phys.~Rev.~E}, \textit{66}, 026,303.

\bibitem[{\textit{Drury}(1987)}]{Drury87}
Drury, M. (1987), Thermal diffusivity of some crystalline rocks,
  \textit{Geothermics}, \textit{16}(2), 105--115.

\bibitem[{\textit{Ebner et~al.}(2010)\textit{Ebner, Toussaint, Schmittbuhl,
  Koehn, and Bons}}]{Ebner10}
Ebner, M., R.~Toussaint, J.~Schmittbuhl, D.~Koehn, and P.~Bons (2010),
  Anisotropic scaling of tectonic stylolites: a fossilized signature of the
  stress field?, \textit{J.~Geophys.~Res.}, \textit{115}, B06,403,
  \doi{10.1029/2009JB006649}.

\bibitem[{\textit{Flekk{\o}y}(1993)}]{FlekkoyBGK93}
Flekk{\o}y, E. (1993), Lattice {B}hatnagar-{G}ross-{K}rook models for miscible
  fluids, \textit{Phys.~Rev.~E}, \textit{47}(6), 4247--4257.

\bibitem[{\textit{Gelet et~al.}(2012)\textit{Gelet, Loret, and
  Khalili}}]{Gelet12}
Gelet, R., B.~Loret, and N.~Khalili (2012), A thermo-hydro-mechanical coupled
  model in local thermal non-equilibrium for fractured {HDR} reservoir with
  double porosity, \textit{J.~Geophys.~Res.}, \textit{117}(7).

\bibitem[{\textit{Grebenkov et~al.}(2007)\textit{Grebenkov, Filoche, and
  Sapoval}}]{Grebenkov07}
Grebenkov, D., M.~Filoche, and B.~Sapoval (2007), A simplified analytical model
  for laplacian transfer across deterministic prefractal interfaces,
  \textit{Fractals}, \textit{15}(1), 27--39.

\bibitem[{\textit{Guyon et~al.}(2001)\textit{Guyon, Hulin, and
  Petit}}]{GuyonHulinPetit}
Guyon, E., J.-P. Hulin, and L.~Petit (2001), \textit{Hydrodynamique physique},
  EDP Sciences.

\bibitem[{\textit{Halecky et~al.}(2011)\textit{Halecky, Birkholzer, and
  Peterson}}]{Halecky11}
Halecky, N., J.~Birkholzer, and P.~Peterson (2011), Natural convection in
  tunnels at yucca mountain and impact on drift seepage, \textit{Nuclear
  Technology}, \textit{174}(3), 327--341.



\bibitem[{\textit{Harting et~al.}(2010)\textit{Harting, Kunert, and
      Hyväluoma}}]{Harting10} Harting, J. and C.~Kunert, and
  J.~Hyväluoma (2010), Lattice {B}oltzmann simulations in
  microfluidics: probing the no-slip boundary condition in
  hydrophobic, rough, and surface nanobubble laden microchannels,
  \textit{Microfluidics and Nanofluidics}, \textit{8}(1), 1613--4982.


\bibitem[{\textit{He and Luo}(1997)\textit{He and Luo}}]{HeLuo97}
  X.~He and L.S. Luo. (1997), Lattice {B}oltzmann model for the
  incompressible {N}avier-{S}tokes equation. 
  \emph{J. Stat. Phys.}, 88:\penalty0 927--944, 1997.


\bibitem[{\textit{Hiorth et~al.}(2008)\textit{Hiorth, {a Lad}, Evje, and
  S.M.}}]{Hiorth08}
Hiorth, A., U.~{a Lad}, S.~Evje, and S.~S.M. (2008), A lattice
  {B}oltzmann-{BGK} algorithm for a diffusion equation with {R}obin boundary
  condition—application to {NMR} relaxation, \textit{Int. J. Numer. Meth.
  Fluids}, \textit{59}(4), 405--421, \doi{10.1002/fld.1822}.

\bibitem[{\textit{Hoteit et~al.}(2004)\textit{Hoteit, Ackerer, and
  Mosé}}]{Hoteit04}
Hoteit, H., P.~Ackerer, and R.~Mosé (2004), Nuclear waste disposal
  simulations: {C}ouplex test cases, \textit{Computational Geosciences},
  \textit{8}(2), 99--124.

\bibitem[{\textit{Huang et~al.}(2011)\textit{Huang, Lu, and Sukop}}]{Huang11}
Huang, H.-B., X.-Y. Lu, and M.~Sukop (2011), Numerical study of lattice
  {B}oltzmann methods for a convection-diffusion equation coupled with
  {N}avier-{S}tokes equations, \textit{Journal of Physics A: Mathematical and
  Theoretical}, \textit{44}(5).

\bibitem[{\textit{Johnsen et~al.}(2006)\textit{Johnsen, Toussaint,
  M\aa{}l\o{}y, and Flekk\o{}y}}]{Johnsen06}
Johnsen, O., R.~Toussaint, K.~M\aa{}l\o{}y, and E.~Flekk\o{}y (2006), Pattern
  formation during air injection into granular materials confined in a circular
  hele-shaw cell, \textit{Phys.~Rev.~E}, \textit{74}, 011,301,
  \doi{10.1103/PhysRevE.74.011301}.

\bibitem[{\textit{Kobchenko et~al.}(2011)\textit{Kobchenko, Panahi, Renard,
  Dysthe, Malthe-Srenssen, Mazzini, Scheibert, Jamtveit, and
  Meakin}}]{Kobchenko11}
Kobchenko, M., H.~Panahi, F.~Renard, D.~Dysthe, A.~Malthe-Srenssen, A.~Mazzini,
  J.~Scheibert, B.~Jamtveit, and P.~Meakin (2011), 4{D} imaging of fracturing
  in organic-rich shales during heating, \textit{J.~Geophys.~Res.},
  \textit{116}(12).

\bibitem[{\textit{Koehn et~al.}(2012)\textit{Koehn, Ebner, Renard, Toussaint,
  and Passchier}}]{Koehn12}
Koehn, D., M.~Ebner, F.~Renard, R.~Toussaint, and C.~Passchier (2012),
  Modelling of stylolite geometries and stress scaling, \textit{Earth and
  Planetary Science Letters}, \textit{341-344}, 104--113.

\bibitem[{\textit{Lallemand and Luo}(2003)}]{Lallemand03}
Lallemand, P., and L.~Luo (2003), Theory of the lattice boltzmann method:
  Acoustic and thermal properties in two and three dimensions,
  \textit{Phys.~Rev.~E}, \textit{68}.

\bibitem[{\textit{Lan et~al.}(2012)\textit{Lan, Martin, and Andersson}}]{Lan12}
Lan, H., C.~Martin, and J.~Andersson (2012), Evolution of in situ rock mass
  damage induced by mechanical-thermal loading, \textit{Rock Mech. Rock Eng.},
  pp. 1--16, article in Press.

\bibitem[{\textit{Laronne Ben-Itzhak et~al.}(2012)\textit{Laronne Ben-Itzhak,
  Aharonov, Toussaint, and Sagy}}]{Laronne12}
Laronne Ben-Itzhak, L., E.~Aharonov, R.~Toussaint, and A.~Sagy (2012), Upper
  bound on stylolite roughness as indicator for amount of dissolution,
  \textit{Earth Planet. Sci. Lett.}, \textit{337-338}, 186--196.

\bibitem[{\textit{Luan et~al.}(2012)\textit{Luan, Xu, Chen, Feng, He, and
  Tao}}]{Luan12}
Luan, H., H.~Xu, L.~Chen, Y.~Feng, Y.~He, and W.~Tao (2012), Coupling of finite
  volume method and thermal lattice boltzmann method and its application to
  natural convection, \textit{International Journal for Numerical Methods in
  Fluids}, \textit{70}(2), 200--221.

\bibitem[{\textit{Moffatt}(1964)}]{Moffatt64}
Moffatt, H. (1964), Viscous and resistive eddies near a sharp corner,
  \textit{J. Fluid Mech.}, \textit{18}, 1--18.

\bibitem[{\textit{Mollo et~al.}(2011)\textit{Mollo, Vinciguerra, Iezzi,
  Iarocci, Scarlato, Heap, and Dingwell}}]{Mollo11}
Mollo, S., S.~Vinciguerra, G.~Iezzi, A.~Iarocci, P.~Scarlato, M.~Heap, and
  D.~Dingwell (2011), Volcanic edifice weakening via devolatilization
  reactions, \textit{Geophys. J. Int.}, \textit{186}(3), 1073--1077.

\bibitem[{\textit{Natarajan and Kumar}(2010)}]{Natarajan_thermal_10}
Natarajan, N., and G.~Kumar (2010), Thermal transport in a coupled sinusoidal
  fracture-matrix system, \textit{International Journal of Engineering Science
  and Technology}, \textit{2,}(7), 2645--2650.

\bibitem[{\textit{Nenna and Aydin}(2011)}]{Nenna11}
Nenna, F., and A.~Aydin (2011), The role of pressure solution seam and joint
  assemblages in the formation of strike-slip and thrust faults in a
  compressive tectonic setting; {T}he {V}ariscan of south-western {I}reland,
  \textit{J. Struct. Geol.}, \textit{33}(11), 1595--1610.

\bibitem[{\textit{Neuville et~al.}(2010)\textit{Neuville, Toussaint, and
  Schmittbuhl}}]{Neuville10PRE}
Neuville, A., R.~Toussaint, and J.~Schmittbuhl (2010), Hydro-thermal flows in a
  self-affine rough fracture, \textit{Phys.~Rev.~E}, \textit{82}, 036,317,
  \doi{10.1103/PhysRevE.82.036317}.

\bibitem[{\textit{Neuville et~al.}(2011)\textit{Neuville, Toussaint, and
  Schmittbuhl}}]{Neuville10JGI}
Neuville, A., R.~Toussaint, and J.~Schmittbuhl (2011), Hydraulic transmissivity
  and heat exchange efficiency of open fractures: a model based on lowpass
  filtered apertures., \textit{Geophys. J. Int.}, \textit{186}(3), 1064--1072,
  \doi{10.1111/j.1365-246X.2011.05126.x}.

\bibitem[{\textit{Neuville et~al.}(2012)\textit{Neuville, Toussaint,
  Schmittbuhl, Koehn, and Schwarz}}]{NeuvilleDraixI09}
Neuville, A., R.~Toussaint, J.~Schmittbuhl, D.~Koehn, and J.~Schwarz (2012),
  Characterization of major discontinuities from borehole cores of the black
  consolidated marl formation of {D}raix ({F}rench {A}lps), \textit{Hydrol.
  Processes}, \textit{26}(14), 2095–2105, \doi{"10.1002/hyp.7984"}, special
  issue, hydrology of clay shales and clayey sediments.


\bibitem[\textit{Nicholl et~al.}(1999)\textit{Nicholl, M., Rajaram, H., Glass, R., and Detwiler, R.}]{Nicholl99}
Nicholl, M., Rajaram, H., Glass, R., and Detwiler, R. (1999).
\newblock Saturated flow in a single fracture: Evaluation of the {R}eynolds
  equation in measured aperture fields.
\newblock {\textit Water Resour. Res.}, 35(11):3361--3373.


\bibitem[{\textit{Niebling et~al.}(2010)\textit{Niebling, Flekk\o{}y,
  M\aa{}l\o{}y, and Toussaint}}]{Niebling10}
Niebling, M., E.~Flekk\o{}y, K.~M\aa{}l\o{}y, and R.~Toussaint (2010),
  Sedimentation instabilities: {I}mpact of the fluid compressibility and
  viscosity, \textit{Phys.~Rev.~E}, \textit{82}, 051,302,
  \doi{10.1103/PhysRevE.82.051302}.

\bibitem[\textit{Oron and Berkowitz}(1998)]{Oron98}
Oron, A. and Berkowitz, B. (1998).
\newblock Flow in rock fractures: The local cubic law assumption reexamined.
\newblock {\textit{Water Resour. Res.}}, \textit{34}(11):2811--2825.

\bibitem[{\textit{Patankar}(1980)}]{Patankar80}
Patankar, S. (1980), \textit{Numerical Heat Transfer and Fluid Flow},
  Hemisphere Publishing Corporation, New York.

\bibitem[{\textit{Petersen et~al.}(2010)\textit{Petersen, Bojesen-Koefoed, and
  Mathiesen}}]{Petersen10}
Petersen, H., J.~Bojesen-Koefoed, and A.~Mathiesen (2010), Variations in
  composition, petroleum potential and kinetics of ordovician - miocene {T}ype
  {I} and {T}ype {I}-{II} source rocks (oil shales): {I}mplications for
  hydrocarbon generation characteristics, \textit{Journal of Petroleum
  Geology}, \textit{33}(1), 19--41.

\bibitem[{\textit{Pinkus and Sternlicht}(1961)}]{Pinkus61}
Pinkus, O., and B.~Sternlicht (1961), \textit{Theory of Hydrodynamic
  Lubrication}, McGraw-Hill, {New York, U.S.A}, 465~p.

\bibitem[{\textit{Qian et~al.}(1992)\textit{Qian, d'Humières, and
  Lallemand}}]{Qian92}
Qian, Y., D.~d'Humières, and P.~Lallemand (1992), Lattice {BGK} models for
  {N}avier-{S}tokes equation, \textit{Europhys. Lett.}, \textit{17}(6), 479.

\bibitem[{\textit{Renard et~al.}(2004)\textit{Renard, Schmittbuhl, Gratier,
  Meakin, and Merino}}]{Renard04}
Renard, F., J.~Schmittbuhl, J.-P. Gratier, P.~Meakin, and E.~Merino (2004),
  Three-dimensional roughness of stylolites in limestones,
  \textit{J.~Geophys.~Res.}, \textit{109}(3).

\bibitem[{\textit{Renard et~al.}(2009)\textit{Renard, Bernard, Desrues, and
  Ougier-Simonin}}]{RenardBernard09}
Renard, F., D.~Bernard, J.~Desrues, and A.~Ougier-Simonin (2009), 3{D} imaging
  of fracture propagation using synchrotron {X}-ray microtomography,
  \textit{Earth Planet. Sci. Lett.}, \textit{286}(1-2), 285--291.

\bibitem[{\textit{Rolland et~al.}(2012)\textit{Rolland, Toussaint, Baud,
  Schmittbuhl, Conil, Koehn, Renard, and Gratier}}]{Rolland12}
Rolland, A., R.~Toussaint, P.~Baud, J.~Schmittbuhl, N.~Conil, D.~Koehn,
  F.~Renard, and J.-P. Gratier (2012), Modeling the growth of stylolites in
  sedimentary rocks, \textit{J.~Geophys.~Res.}, \textit{117}(6).

\bibitem[{\textit{Rothman and Zaleski}(1997)}]{RothZaleski}
Rothman, D., and S.~Zaleski (1997), \textit{Lattice {B}oltzmann Methods for
  Flow in Porous Media}, Cambridge University Press.

\bibitem[{\textit{Steefel et~al.}(2005)\textit{Steefel, DePaolo, and
  Lichtner}}]{Steefel05}
Steefel, C., D.~DePaolo, and P.~Lichtner (2005), Reactive transport modeling:
  an essential tool and a new research approach fore the earth sciences,
  \textit{Earth and Planetary Science Letters}, \textit{240}, 539--558,
  \doi{10.1016/j.espl.2005.09.17}.

\bibitem[{\textit{Stephansson et~al.}(2004)\textit{Stephansson, Hudson, and
  Lanru}}]{Stephansson04}
Stephansson, O., J.~Hudson, and J.~Lanru (2004), \textit{Coupled
  Thermo-Hydro-Mechanical-Chemical Processes in Geo-systems}, vol.~2, Elsevier
  Science.

\bibitem[{\textit{Szymczak and Ladd}(2009)}]{Szymczak09}
Szymczak, P., and A.~Ladd (2009), Wormhole formation in dissolving fractures,
  \textit{J. Geophys. Res., B}, \textit{114}(6).

\bibitem[{\textit{Taine and Petit}(2003)}]{Taine03}
Taine, J., and J.-P. Petit (2003), \textit{Transferts Thermiques},
  $3^\text{rd}$ ed., Dunod, Paris, {F}rance, 449~p.

\bibitem[{\textit{Talon et~al.}(2012)\textit{Talon, Bauer, Gland, Youssef,
  Auradou, and Ginzburg}}]{Talon12}
Talon, L., D.~Bauer, N.~Gland, S.~Youssef, H.~Auradou, and I.~Ginzburg (2012),
  Assessment of the two relaxation time {L}attice-{B}oltzmann scheme to
  simulate stokes flow in porous media, \textit{Water Resour. Res.},
  \textit{48}(4).

\bibitem[{\textit{Turcotte and Schubert}(2002)}]{TurcotteSchu}
Turcotte, D., and G.~Schubert (2002), \textit{Geodynamics}, chap.~6, pp.
  262--264, $2^\text{nd}$ ed., Cambridge University Press.

\bibitem[{\textit{Vinningland et~al.}(2012)\textit{Vinningland, Toussaint,
  Niebling, Flekk\o{}y, and M\aa{}l\o{}y}}]{Vinningland12}
Vinningland, J., R.~Toussaint, M.~Niebling, E.~Flekk\o{}y, and K.~M\aa{}l\o{}y
  (2012), Family-{V}icsek scaling of detachment fronts in granular
  {R}ayleigh-{T}aylor instabilities during sedimentating granular/fluid flows,
  \textit{European Physical Journal: Special Topics}, \textit{204}(1), 27--40.

\bibitem[{\textit{Waite et~al.}(1998)\textit{Waite, Ge, and
  Spetzler}}]{Miles98}
Waite, M., S.~Ge, and H.~Spetzler (1998), The effect of surface geometry on
  fracture permeability: A case study using a sinusoidal fracture,
  \textit{Geophys.~Res.~Lett.}, \textit{25}(6), 813--816.

\bibitem[{\textit{Wolf-Gladrow}(2005)}]{WolfG05}
Wolf-Gladrow, D. (2005), \textit{Lattice-Gas Cellular Automata and Lattice
  {B}oltzmann Models -- {A}n Introduction}, Springer.

\bibitem[{\textit{Yeo}(2001)}]{Yeo11}
Yeo, I. (2001), Effect of fracture roughness on solute transport,
  \textit{Geosciences Journal}, \textit{5}(2), 145--151.

\bibitem[{\textit{Young}(1998)}]{Young98}
Young, T., and K.~Vafai (1998), Convective cooling of a heated
obstacle in a channel, \textit{International Journal of Heat and Mass Transfer}
  \textit{41}(20), 3131--3148.


\bibitem[{\textit{Zimmerman and Bodvarsson}(1996)}]{Zimmerman96}
Zimmerman, R., and G.~Bodvarsson (1996), Hydraulic conductivity of rock
  fractures, \textit{Transp. Porous Media}, \textit{23}(1), 1--30.

\end{thebibliography}
%
%
%
%
%
%




%
%

\end{article}

s and Tables here:



%
%
%
%
%
%


\end{document}